\documentclass[11pt,notitlepage,nofootinbib]{revtex4-1}

\usepackage{epsf}
\usepackage{latexsym}
\usepackage{amssymb}
\usepackage{amsmath}
\usepackage{bm}
\usepackage[dvips]{graphicx}
\usepackage{array}

\usepackage{color}

\date{\today}
\def\d3{^{(3)}\nabla}

\begin{document}

\title{CMB anisotropies and linear matter power spectrum in models with non-thermal neutrinos and primordial magnetic fields}

\author{Kerstin E. Kunze}

\email{kkunze@usal.es}

\affiliation{Departamento de F\'\i sica Fundamental, Universidad de Salamanca,
 Plaza de la Merced s/n, 37008 Salamanca, Spain}

\begin{abstract}
Angular power spectra of temperature anisotropies and polarization of the cosmic microwave background (CMB) 
as well as the linear matter power spectra are calculated for models with three light neutrinos with
non-thermal phase-space distributions in the presence of a primordial stochastic magnetic field.
The non-thermal phase-space distribution function is assumed to be the sum of a Fermi-Dirac  and a  gaussian distribution. 
It is found that the known effective description of the non-thermal model in terms of a twin thermal model with extra relativistic degrees of freedom 
can also be extended to models including a stochastic magnetic field.
Numerical solutions are obtained for a range of magnetic field parameters.

\end{abstract}

\maketitle

\section{Introduction}
\label{s0}
\setcounter{equation}{0}

The standard model of cosmology predicts the presence of a cosmic neutrino background.
Direct detection is very difficult. Neutrinos in the standard model are light and have a thermal distribution.
Observations of solar as well as atmospheric neutrino oscillations put bounds on the mass differences.
Depending on which mass difference is assigned between the two lowest neutrino masses the mass 
hierarchy is either normal or inverted. Effectively, this results in either two lighter neutrinos and one more
massive or vice versa.
Distinguishing between the mass hierarchies is very challenging as
the mass differences are extremely small. However, in principle it might be possible using cosmological observations as neutrinos with different masses have slightly different decoupling temperatures.
Light neutrinos act as non cold dark matter. The main effect is the suppression of the linear matter power spectrum on small scales due to free-streaming
(for reviews, e.g., \cite{pdg2020, lmmp,Hannestad:2010kz,Giunti:2007ry,Dolgov:2002wy}).

Primordial magnetic fields have the opposite effect on the matter power spectrum. The Lorentz term in the 
baryon velocity equation drives the  evolution of the baryon and consequently the dark matter perturbation
causing the domination over the adiabatic, primordial curvature mode on small scales \cite{sl,SeSu,KimOlinRos, wasserman}.
Primordial magnetic fields generated in the very early universe, e.g. during inflation or at a phase transition,
are generally constrained to be less or of order nG, in terms of their present day comoving field strength (e.g., \cite{planck15-mag}).
Taking into account the evolution of cosmic magnetic fields due to their interaction with the cosmic plasma, leading to 
damping by decaying magnetohydrodynamic turbulence shortly after photon decoupling or ambipolar diffusion important in 
partially ionized matter, before the epoch of reionization, much stronger constraints have been derived.
This is due to the additional heating of matter by the dissipating magnetic field \cite{CPF,kuko15,SeSu}

Within the standard model of cosmology neutrinos are in thermal equilibrium with the rest of the cosmic plasma 
upto their decoupling when the universe cooled down to about 1 MeV  resulting in the postulated cosmic neutrino background (C$\nu$B). The Fermi-Dirac distribution could include a non zero chemical potential which, however, can be constrained by Big Bang Nucleosynthesis (BBN).
Neutrinos with a non-thermal distribution could be generated, e.g., 
in particle physics models beyond the standard model with very massive nonrelativistic unstable particles  whose decay can lead to late time entropy production
and reheat temperatures of ${\cal O}$(1 MeV). Different decay scenarios of the  massive particles and neutrino production have been studied  \cite{Kawasaki:1999na,Kawasaki:2000en} and also together with the effects of neutrino oscillation as well as neutrino self-interaction in the neutrino thermalization process \cite{Hasegawa:2019jsa}.
Neutrinos with a non-thermal distribution could also be created by particle decay at a much later stage. 
Modelling this non-thermal contribution as a gaussian it has been shown in \cite{Cuoco:2005} that the non-thermal model can be described in terms of a "twin" model with thermal neutrinos and extra (relativistic) degrees of freedom.
Thereby allowing for a degeneracy between this type of non-thermal C$\nu$B and a thermal counterpart in a cosmic background with extra relativistic degrees of freedom.
It is this model of non-thermal neutrinos which will be used here in backgrounds with a primordial stochastic magnetic field.

There are several open source programs available to calculate the CMB anisotropies and the matter power spectrum, such as COSMICS
\cite{Bertschinger:1995er}, CMBFAST \cite{Seljak:1996is}, CAMB \cite{Lewis:1999bs}, CMBEASY \cite{Doran:2003sy} 
and CLASS \cite{class1,class2,class3,class4,class5}. 
The numerical solutions are obtained here by modifying the CLASS code.

In section \ref{s1} details of the non-thermal neutrino phase-space distribution function and the contribution of the primordial stochastic magnetic field
are provided. Results are  given in section \ref{s2}. Section \ref{s3} contains the conclusions.

\section{Modelling non-thermal neutrinos and primordial magnetic fields}
\label{s1}
\setcounter{equation}{0}

In this section details of the model under consideration will be given.

\subsection{Non-thermal neutrinos}

Within the standard model of cosmology BBN constrains the thermal evolution of the universe and predicts the creation of the thermal C$\nu$B at around 1 MeV.
Following \cite{Cuoco:2005} it is assumed that a non-thermal contribution to the C$\nu$B is created by the decay of a light neutral scalar particle $\Phi$ with mass $M$ producing active neutrinos of the same type as in the standard model after the weak interaction freeze-out and before photon decoupling. The latter ensures that the CMB anisotropies and the linear matter power spectrum  are directly affected. The simplest model considers an out-of-equilibrium, instantaneous decay scenario $\Phi\rightarrow \bar{\nu}\nu$ taking place when the C$\nu$B  is at a temperature $T_D$. Upto that moment the neutrinos of the C$\nu$B are determined by a Fermi-Dirac distribution function  with zero chemical potential in the simplest model of standard cosmology. 
With the sudden decay of the $\Phi$ particle and production of neutrinos  the neutrino distribution function receives a non-thermal contribution. 
In \cite{Cuoco:2005} this is modelled by a gaussian distribution function.
In particular, it is assumed that for each mass eigenstate the total neutrino distribution function $f(y)$ is given by 
\begin{eqnarray}
y^2f(y)=y^2\frac{1}{e^y+1}dy+y^2\frac{A}{\sqrt{2\pi\sigma^2}}
\exp\left[-\left(\frac{y-y_*}{2\sigma^2}\right)^2\right],
\label{pdf}
\end{eqnarray}
where the $y=ka$ is the comoving neutrino momentum. The first term is the Fermi-Dirac distribution with zero chemical potential for the thermal part and the second one the
non-thermal distribution function which is strongly peaked at $y_*$ given by 
$y_*=\frac{M}{2T_D}$ and $\sigma\ll y_*$. 

In general the moments of a phase-space distribution can be used to calculate physical variables. Following  \cite{Cuoco:2005} the moments for a neutrino mass state 
$\alpha$ are defined by
\begin{eqnarray}
Q^{(n)}_{\alpha}=\frac{1}{\pi^2}\left(\frac{4}{11}\right)^{\frac{3+n}{3}}T_{\gamma}^{3+n}
\int y^{2+n}f_{\alpha}(y)dy
\end{eqnarray}
where $T_{\gamma}$ is the photon temperature.
For $\alpha$ neutrino mass states the effective number of neutrinos, $N_{eff}$ is given by  \cite{Cuoco:2005}
\begin{eqnarray}
N_{eff}=\frac{120}{7\pi^2}\left(\frac{11}{4}\right)^{\frac{4}{3}}
T_{\gamma}^{-4}\sum_{\alpha}Q_{\alpha}^{(1)},
\end{eqnarray}
Assuming that all neutrino mass states have the same phase-space distribution function
the neutrino density parameter
$\omega_{\nu}=\Omega_{\nu}h^2$,
can be expressed as, suppressing the index $\alpha$,
\begin{eqnarray}
\omega_{\nu}&=&\frac{m_{0}}{94.1[93.2] {\rm eV}}
\frac{2\pi^2}{3\zeta(3)}
\frac{11}{4}
T_{\gamma}^{-3} Q^{(0)}
\end{eqnarray}
where the total mass $m_0$ is written in terms of the standard value for thermal neutrinos with a 
Fermi-Dirac distribution with zero chemical potential for, respectively,  $N_{eff}=3$ and the number in brackets 
for $N_{eff}=3.04$ resulting from more precise numerical solutions including heating during the
electron-positron annihilation phase \cite{pdg2020,Cuoco:2005}. 
As pointed out in  \cite{Cuoco:2005}  cosmological data provide the strongest constraints on 
$N_{eff}$ and $\omega_{\nu}$.
Therefore only the first two moments will be taken into account.
A positive amplitude $A$ in the total neutrino distribution function increases $N_{eff}$ as well as $\omega_{\nu}$.
However, adjusting different parameters 
in the model allows to find correspondences with different kind of cosmological models.
Namely, the observational implications of the three non-thermal neutrino model under consideration here can  
effectively be obtained from a three thermal neutrino model 
with extra relativistic degrees  of freedom  \cite{Cuoco:2005,crotty,lmmp}.

In particular, adjusting the present day value of the neutrino temperature, the neutrino masses as well as the cold dark matter 
density parameter, $\omega_c$,  allows to tune $N_{eff}$, $\omega_{\nu}$ as well as the redshift of radiation-matter equality, $z_{eq}$.
The first  two changes yield the  three thermal neutrino model corresponding to  the non-thermal one. 
A larger $N_{eff}$ leads to a potentially significant change in $z_{eq}$ and hence the amplitude of the first peaks  in the angular power spectrum of the CMB anisotropies.
This can be prevented by adjusting $\omega_c$ accordingly  \cite{bashSelj,Hou:2011ec,lmmp,Follin:2015hya}.

\subsection{Primordial magnetic fields}

Primordial magnetic fields present from before decoupling have a direct influence on the CMB anisotropies
as well as the matter power spectrum determining 
large scale structure (LSS) (e.g. \cite{kb,sl,kk11,pfp}).
This is due to the magnetic energy density perturbation with amplitude in Fourier space, $\Delta_B(\vec{k})$, as well as the anisotropic stress perturbation, $\pi_B(\vec{k})$. Moreover, the Lorentz term with amplitude in Fourier space, 
\begin{eqnarray}
L(\vec{k})=\Delta_B(\vec{k})-\frac{2}{3}\pi_B(\vec{k}), 
\label{L}
\end{eqnarray}
in the evolution equation of the baryon velocity 
causes the rise in the matter power spectrum on small scales.
Assuming a nonhelical, random Gaussian magnetic field its two-point function in $k$-space is chosen to be (e.g., \cite{kk11})
\begin{eqnarray}
\langle B_i^*(\vec{k})B_j(\vec{q})\rangle=(2\pi)^3\delta({\vec{k}-\vec{q}})P_B(k)\left(\delta_{ij}-\frac{k_ik_j}{k^2}\right),
\end{eqnarray}
where the power spectrum, $P_B(k)$, is assumed to be a power law,
$P_B(k)=A_Bk^{n_B}$, with amplitude, $A_B$, and spectral index, $n_B$.
The ensemble average energy density of the magnetic field is defined using a Gaussian window function so that
\begin{eqnarray}
\langle\rho_{B,0}\rangle=\int\frac{d^3k}{(2\pi)^3}P_{B,0}(k)e^{-2\left(\frac{k}{k_c}\right)^2}, 
\label{rhoB0}
\end{eqnarray}
where $k_c$ is a certain Gaussian smoothing scale and a "0" refers to the present epoch. 
The magnetic field is treated as frozen-in with the cosmic fluid. This implies that its energy density scales with the scale factor $a$ as $a^{-4}$.
Effectively, this is implemented by defining the Fourier transformations of the magnetic energy density and anisotropic stress in terms of the photon energy density, yielding  time independent Fourier amplitudes, $\Delta_B(\vec{k})$ and $\pi_B(\vec{k})$, respectively.
Moreover, the magnetic field power spectrum can be conveniently expressed in terms of $\rho_{B,0}$ resulting in
\begin{eqnarray}
P_{B,0}(k)=\frac{4\pi^2}{k_c^3}\frac{2^{(n_B+3)/2}}{\Gamma\left(\frac{n_B+3}{2}\right)}\left(\frac{k}{k_c}\right)^{n_B}\langle\rho_{B,0}\rangle.
\end{eqnarray}
The limit $n_B\rightarrow -3$ approaches the  scale-invariant case 
for which the contribution to the energy density per logarithmic wavenumber
is independent of wave number. 
In general the spectral index of the magnetic field depends on the generation mechanism. A large class of models generate primordial magnetic fields long before recombination. Basically there are two different classes. Firstly magnetic fields generated during inflation amplifying perturbations in the electromagnetic field on superhorizon scales, similar to the generation of the primordial curvature perturbation. These models result in negative spectral indices. Secondly using Biermann battery type mechanisms during a phase transition generate magnetic fields with positive spectral indices. Whereas there is no problem for inflationary models to generate magnetic fields with arbitrary correlation length to obtain a sufficiently strong magnetic field to act as a seed field for galactic magnetic fields can be a challenge. The opposite is the case for magnetic fields generated during phase transitions as their correlation length is limited by the horizon scale at the time of creation. Though since these are helical magnetic fields inverse cascade might increase the initial correlation lengths to those necessary  for galactic magnetic fields (for reviews cf., e.g., \cite{Durrer:2013pga,rev4}).
The smallness of the CMB anisotropies are one indication of the global isotropy which rules out a strong homogeneous cosmological magnetic field. Limits on the present day  magnetic field strength are of the order of a few nG. Taking into account different effects of primordial magnetic fields such as  additional heating caused by their dissipation, non-Gaussianity of the CMB or inhomogeneous recombination lead to even stronger bounds (e.g. \cite{Jedamzik:2018itu,CPF,kuko15}). Here the numerical solutions are obtained for magnetic field strengths of the order of nG and negative spectral indices.

Primordial magnetic fields interact with the cosmic plasma. This leads to a damping of the magnetic field. Before decoupling this is due to photon viscosity, a process similar 
to the damping of density perturbations by photon diffusion, i.e. Silk damping.  For a frozen-in magnetic field this is commonly modelled by  a maximal wave number $k_m$ determined by the Alfv\'en velocity and photon diffusion scale at decoupling  \cite{jko,sb}         
\begin{eqnarray}
k_m=\frac{301.45}{\cos\theta}\left(\frac{B_0}{\rm nG}\right)^{-1}{\rm Mpc}^{-1}
\end{eqnarray}
for the Planck 2018 best fit values of the six-parameter base $\Lambda$CDM model from Planck data alone \cite{Planck-2018}. For simplicity the direction cosine of the wave vector and the magnetic field vector $\cos\theta$ is set to 1 \cite{kuko15}.
The Gaussian smoothing scale will be set to the maximal damping wave number, $k_c\equiv k_m$.

Linear cosmological perturbations induced by 
a primordial magnetic field can be separated into two contributions, namely one that is proportional to $\Delta_B(\vec{k})$ and one that is proportional to $\pi_B(\vec{k})$. The total  CMB angular power spectra as well as the linear matter power spectrum are determined by the two-point functions of the corresponding random variables of $\Delta_B(\vec{k})$ 
and $\pi_B(\vec{k})$ and their auto- and cross correlation functions, respectively.
For massless, thermal neutrinos numerical solutions with the magnetic field correlation functions calculated with a Gaussian window function as described above have been obtained with a modified version of CMBEASY for nonhelical \cite{kk11} as well as helical fields for scalar, vector and tensor modes \cite{kk12}. 

In \cite{kk11} the power spectra determining the two-point correlation functions in $k$-space  of the magnetic energy density and the anisotropic  stress in the scalar sector have been calculated. 
These are expressed in terms of the dimensionless power spectrum ${\cal P}_{FG}(k)$ which determines the two-point function
\begin{eqnarray}
\langle F_{\vec{k}}^*G_{\vec{k}'}\rangle=\frac{2\pi^2}{k^3}{\cal P}_{\langle FG \rangle}(k)\delta_{\vec{k},\vec{k}'}
\end{eqnarray}
of two random variables $F$ and $G$.
The autocorrelation function of the magnetic energy density is determined by the dimensionless power spectrum
\begin{eqnarray}
{\cal P}_{\langle\Delta_{\rm B}\Delta_{\rm B}\rangle}(k,k_{\rm m})&=&\frac{1}{\left[\Gamma\left(\frac{n_{\rm B}+3}{2}\right)\right]^2}
\left[\frac{\rho_{\rm B\,0}}{\rho_{\gamma\, 0}}\right]^2\left(\frac{k}{k_{\rm m}}\right)^{2(n_{\rm B}+3)}e^{-\left(\frac{k}{k_{\rm m}}\right)^2}\int_0^{\infty}dz z^{n_{\rm B}+2}e^{-2\left(\frac{k}{k_{\rm m}}\right)^2z^2}\nonumber\\
&&\hspace{1cm}
\int_{-1}^1 dxe^{2\left(\frac{k}{k_{\rm m}}\right)^2zx}\left(1-2zx+z^2\right)^{\frac{n_{\rm B}-2}{2}}\left(1+x^2+2z^2-4zx\right),
\label{pdd}
\end{eqnarray}
where $x\equiv \vec{k}\cdot\vec{q}/kq$ and $z\equiv\frac{q}{k}$.
Moreover, using the average energy density of the magnetic field (cf. equation (\ref{rhoB0})) leads to $\frac{\rho_{B,0}}{\rho_{\gamma,0}}=9.545\times 10^{-8}\left(\frac{B_0}{\rm nG}\right)^2$.

\begin{figure}[h!]
\centerline{\epsfxsize=3.5in\epsfbox{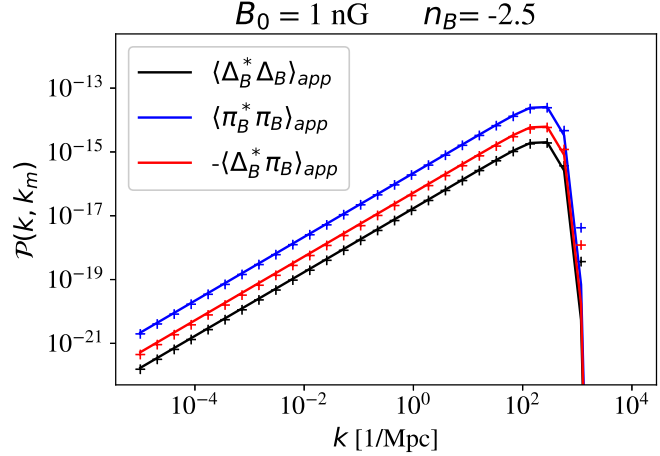}
}
\caption{ Numerical fitting functions (solid lines) $\langle X^*Y\rangle_{app}$ , with $X,Y=\Delta_B, \pi_B$,  together with numerical evaluations (+) of the dimensionless power spectrum of the auto- and cross correlation functions, ${\cal P}_{\langle \Delta_B\Delta_B \rangle}$, ${\cal P}_{\langle \pi_B\pi_B \rangle}$ and $-{\cal P}_{\langle \Delta_B\pi_B \rangle}$ for  magnetic field parameters $B_0=1$ nG, $n_B=-2.5$. }
\label{fig1}
\end{figure}

The numerical solution of ${\cal P}_{\langle\Delta_{\rm B}\Delta_{\rm B}\rangle}(k,k_{\rm m})$ (cf.  equation (\ref{pdd})) can be approximated by the numerical fitting formula
\begin{eqnarray}
{\cal P}_{\langle\Delta_{\rm B}\Delta_{\rm B}\rangle}(k,k_m)&=&\frac{1}{\left[\Gamma\left(\frac{n_{\rm B}+3}{2}\right)\right]^2}
\left[\frac{\rho_{\rm B\,0}}{\rho_{\gamma\, 0}}\right]^2\left(\frac{k}{k_{\rm m}}\right)^{2(n_{\rm B}+3)}e^{-\left(\frac{k}{k_{\rm m}}\right)^2}
\nonumber\\
&&\times 4.09512\left( n_B+3.1\right)^{-1.16608}.
\label{pdd-numfit}
\end{eqnarray}
The autocorrelation function of the magnetic anisotropic stress is determined by the dimensionless spectrum \cite{kk11}
\begin{eqnarray}
{\cal P}_{\langle \pi_{\rm B}\pi_{\rm B} \rangle}(k,k_{\rm m})=\frac{9}{\left[\Gamma\left(\frac{n_{\rm B}+3}{2}\right)\right]^2}
\left[\frac{\rho_{\rm B\,0}}{\rho_{\gamma\, 0}}\right]^2\left(\frac{k}{k_{\rm m}}\right)^{2(n_{\rm B}+3)}e^{-\left(\frac{k}{k_{\rm m}}\right)^2}\int_0^{\infty}dz z^{n_{\rm B}+2}e^{-2\left(\frac{k}{k_{\rm m}}\right)^2z^2}\nonumber\\
\int_{-1}^1 dxe^{2\left(\frac{k}{k_{\rm m}}\right)^2zx}\left(1-2zx+z^2\right)^{\frac{n_{\rm B}-2}{2}}
\left(1+5z^2+2zx+(1-12z^2)x^2
-6zx^3+9z^2x^4\right).
\label{Ppp}
\end{eqnarray}
The numerical solution for ${\cal P}_{\langle \pi_{\rm B}\pi_{\rm B} \rangle}(k,k_{\rm m})$ can be approximated by the numerical fitting formula,
 \begin{eqnarray}
{\cal P}_{\langle \pi_B\pi_B \rangle}(k,k_m)&=&\frac{9}{\left[\Gamma\left(\frac{n_{\rm B}+3}{2}\right)\right]^2}
\left[\frac{\rho_{\rm B\,0}}{\rho_{\gamma\, 0}}\right]^2\left(\frac{k}{k_{\rm m}}\right)^{2(n_{\rm B}+3)}e^{-\left(\frac{k}{k_{\rm m}}\right)^2}
\nonumber\\
&&\times 6.30726 \left( n_B+3.1\right)^{-0.992669}.
\label{Ppp-numfit}
\end{eqnarray}
The cross correlation two-point function of the magnetic energy density and anisotropic stress is determined by the 
dimensionless power spectrum given by 
\begin{eqnarray}
{\cal P}_{\langle\Delta_{\rm B}\pi_{\rm B}\rangle}(k,k_m)=\frac{3}{\left[\Gamma\left(\frac{n_{\rm B}+3}{2}\right)\right]^2}\left[\frac{\rho_{\rm B 0}}{\rho_{\gamma 0}}\right]^2\left(\frac{k}{k_{\rm m}}\right)^{2(n_{\rm B}+3)}e^{-\left(\frac{k}{k_{\rm m}}\right)^2}
\int_0^{\infty}dz z^{n_{\rm B}+2}e^{-2\left(\frac{k}{k_{\rm m}}\right)^2 z^2}
\nonumber\\
\int_{-1}^1dx e^{2\left(\frac{k}{k_{\rm m}}\right)^2zx}
\left(1-2zx+z^2\right)^{ \frac{n_{\rm B}-2}{2}}\left(-1+z^2+zx-(1+3z^2)x^2+3zx^3\right).
\label{Pdp}
\end{eqnarray}
This is well fitted by the numerical fitting formula
\begin{eqnarray}
{\cal P}_{\langle\Delta_{\rm B}\pi_{\rm B}\rangle}(k,k_m)&=&\frac{3}{\left[\Gamma\left(\frac{n_{\rm B}+3}{2}\right)\right]^2}\left[\frac{\rho_{\rm B 0}}{\rho_{\gamma 0}}\right]^2\left(\frac{k}{k_{\rm m}}\right)^{2(n_{\rm B}+3)}e^{-\left(\frac{k}{k_{\rm m}}\right)^2}
\nonumber\\
&&\times
(-3.43671) \left( n_B+3.05\right)^{-0.329634 n_B -2.15842}.
\label{Pdp-numfit}
\end{eqnarray}
The numerical fitting functions together with the full numerical solutions are shown  in figure \ref{fig1}.

Initial conditions correspond to the compensated magnetic mode, an isocurvature type mode,  or the passive magnetic mode, an adiabatic type mode (cf., e.g. \cite{sl}).
The focus here is on the compensated magnetic mode  \cite{kk11}.

\section{Results}
\label{s2}
\setcounter{equation}{0}

The numerical solutions for the CMB angular power spectra as well as the linear matter power spectrum have been obtained by modifying the publicly available Boltzmann solver code
CLASS \cite{class1}-\cite{class5}.  The code has been modified in two ways. Firstly the non-thermal phase space distribution  resulting from an added Gaussian peak  as described in section \ref{s1} (cf. equation (\ref{pdf})) has been included. Secondly the contribution of the magnetic field has been added accordingly to the perturbation equations, the initial conditions for the compensated magnetic mode and the auto- and cross correlation functions for the magnetic energy density perturbation $\Delta_B$ and the anisotropic stress $\pi_B$ using a Gaussian window function as described in section \ref{s1}. For the latter ones the numerical approximations given in equations (\ref{pdd-numfit}), (\ref{Ppp-numfit})
and (\ref{Pdp-numfit}) have been used.
Numerical solutions are obtained for three different models:
\begin{enumerate}
\item TH: 
a model with three thermal neutrinos each of the same mass 
which is chosen to be $m_{\nu}=0.167$ eV, 
\item NT:
a model with three non-thermal neutrinos with the same masses as in the thermal case (model {\it i.)}) and the distribution function (\ref{pdf}) setting the parameters to $A=0.1$, 
$\sigma=1.0$ and $y_c=9.489$,
\item TH+R-twin:
the twin model of the three non-thermal neutrino model ({\it ii.}): 
three thermal neutrinos with extra relativistic degrees of freedom. In this model the neutrino masses are rescaled in order to ensure the same value of the neutrino density parameter 
$\omega_{\nu}$ as in the non-thermal model ({\it ii.}). 
\end{enumerate}
The parameter values of the three non-thermal (NT) neutrino model ({\it ii.)} are chosen as way of example to study in particular the effect on the compensated magnetic mode.
This results in the rather large,  total number of relativistic degrees of freedom, $N^{NT}_{eff}=8.049$. The three thermal neutrino model ({\it i.)} has the standard value, $N^{TH}_{eff}=3.04$.

The cosmological background values are set to $\omega_b=0.022383$, $\omega_c=0.12011$,
$A_s=2.101\times 10^{-9}$ and $\theta_s=1.040909\times 10^{-2}$ which correspond to the Planck 2018 best fit values of the six-parameter base $\Lambda$CDM model from Planck data alone \cite{Planck-2018}. Apart from $\omega_c$ these cosmological 
background values are kept fixed in all numerical solutions. Additional relativistic degrees of freedom as obtained in the three non-thermal neutrino model (cf. 
$N_{eff}^{NT}$  above for the particular choice of parameters for the distribution function equation (\ref{pdf}))  postpone the beginning of the matter dominated epoch  subsequently shortening the time remaining to photon decoupling. This results in larger amplitudes  of the CMB anisotropies. Keeping the baryon density parameter unchanged, which is already constrained quite strongly by LSS, .e.g., \cite{lmmp},  for larger $N_{eff}$ the epoch of radiation-matter equality $z_{eq}$ can be adjusted to $z_{eq}^{TH}$ of the three thermal neutrino model by changing $\omega_c$ accordingly. 
This is indicated in the figures by the additional label $z_{eq}^{TH}$.

\begin{figure}[h!]
\centerline{\epsfxsize=3.2in\epsfbox{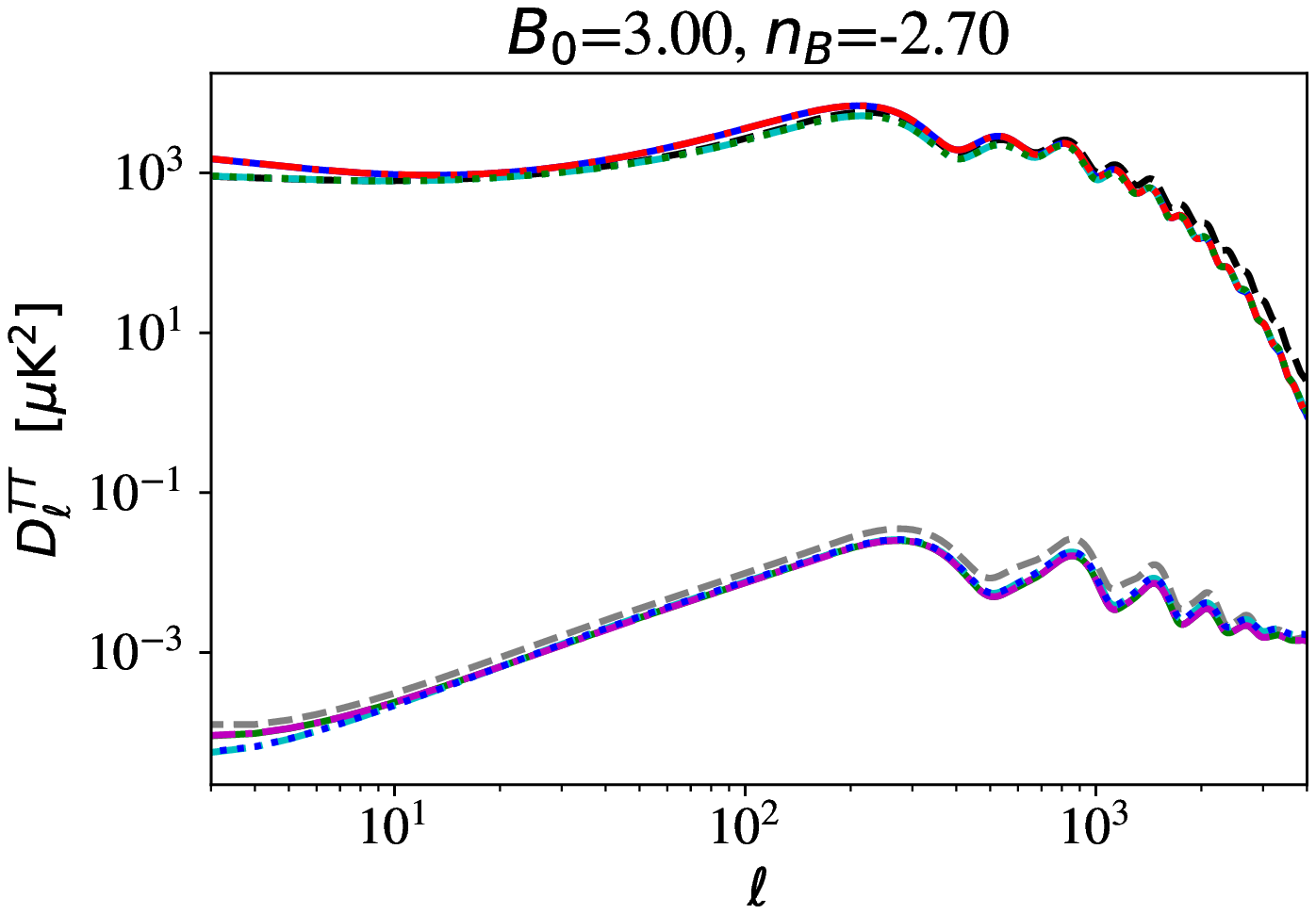}
\hspace{0.05cm}
\epsfxsize=3.2in\epsfbox{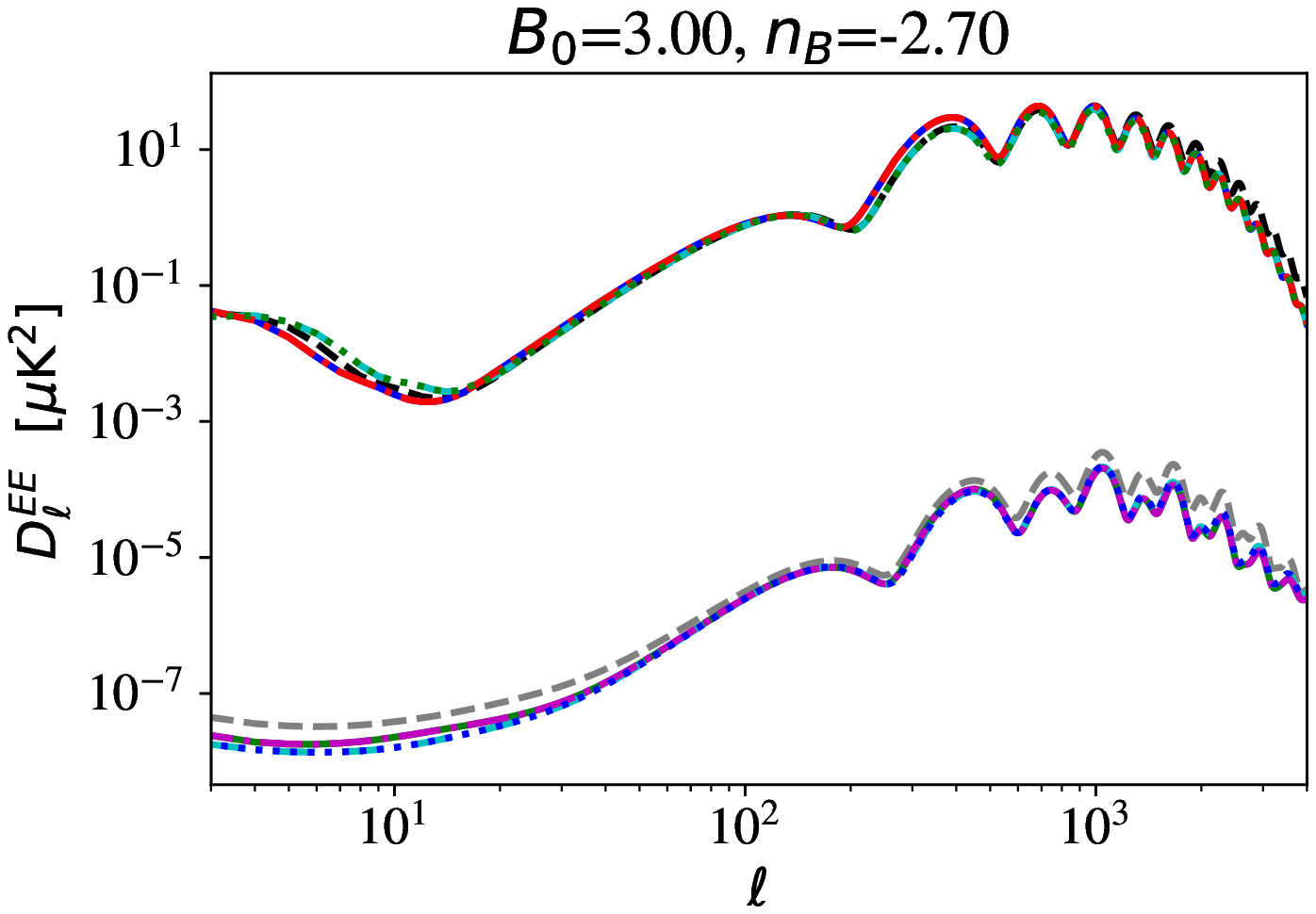}}
\centerline{
\epsfxsize=3.2in\epsfbox{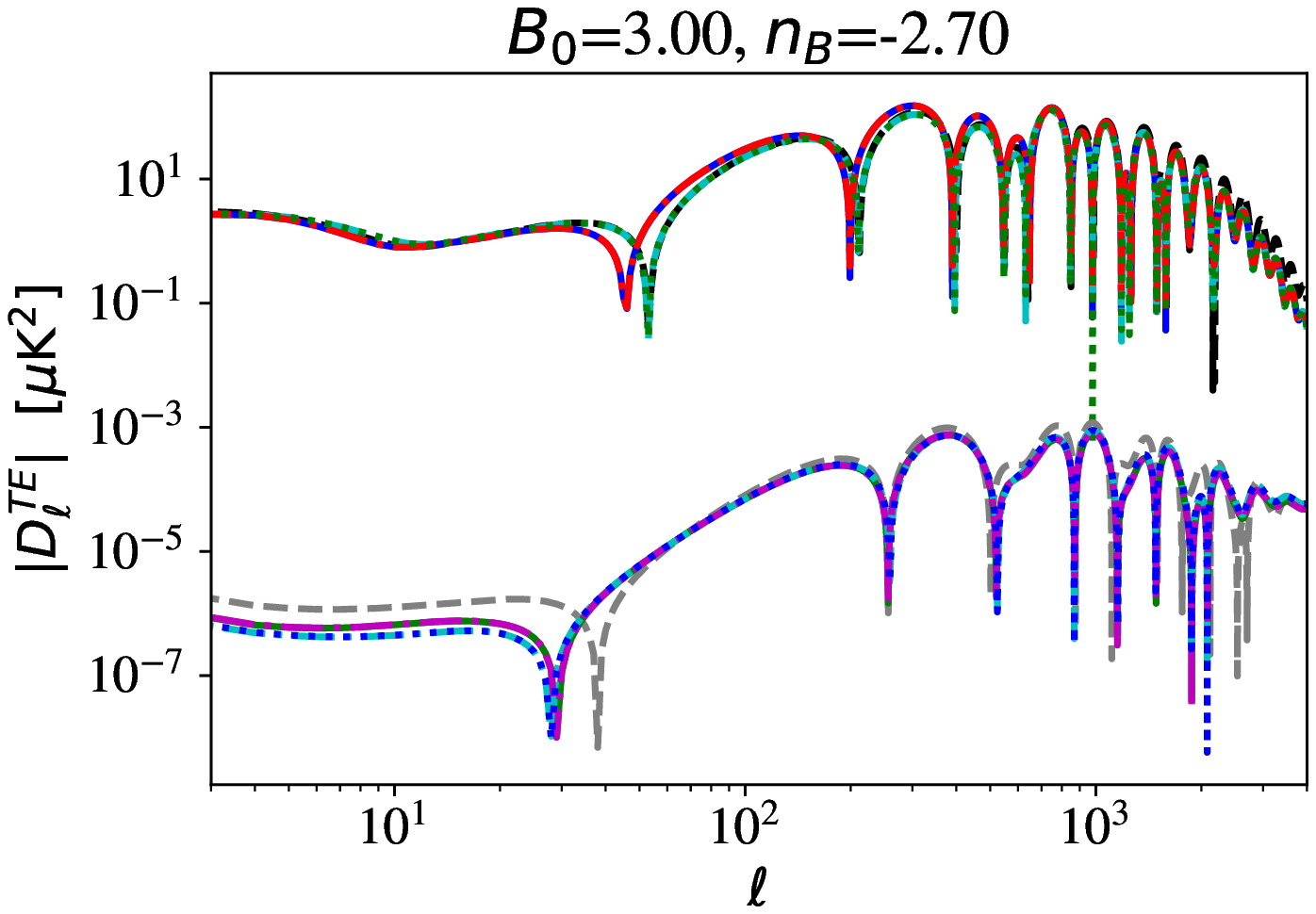}
\hspace{0.5in}
\epsfxsize=2.8in\epsfbox{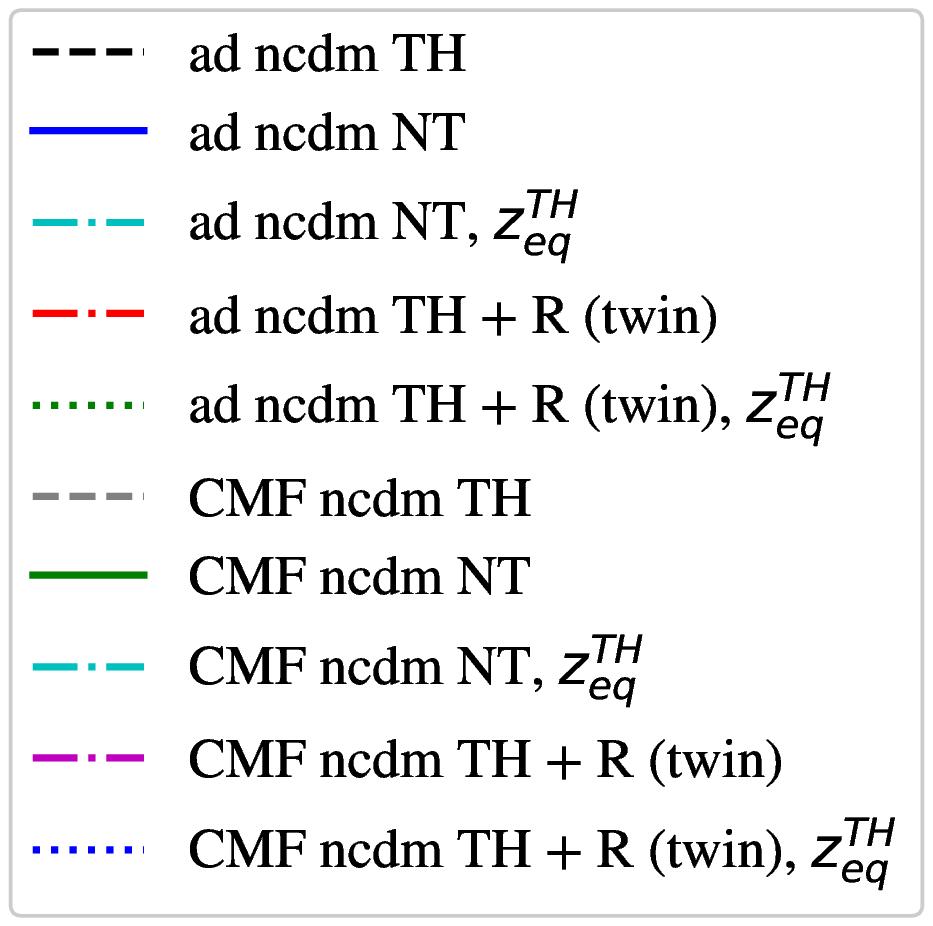}}
\caption{Auto- and cross correlation angular power spectra of the T- and E-mode of models with three thermal neutrinos (TH), three non-thermal neutrinos (NT), three non-thermal neutrinos (NT, $z_{eq}^{TH}$) with $\omega_c$ such that radiation-matter equality occurs at the same redshift as in the three thermal neutrino model $z_{eq}^{TH}$ and their thermal twin models with extra relativistic degrees of freedom 
for the adiabatic mode (ad) and the compensated magnetic mode (CMF) with $B_0=3$ nG, $n_B=-2.70$.
}
\label{fig2}
\end{figure}

\begin{figure}[h!]
\centerline{\epsfxsize=3.2in\epsfbox{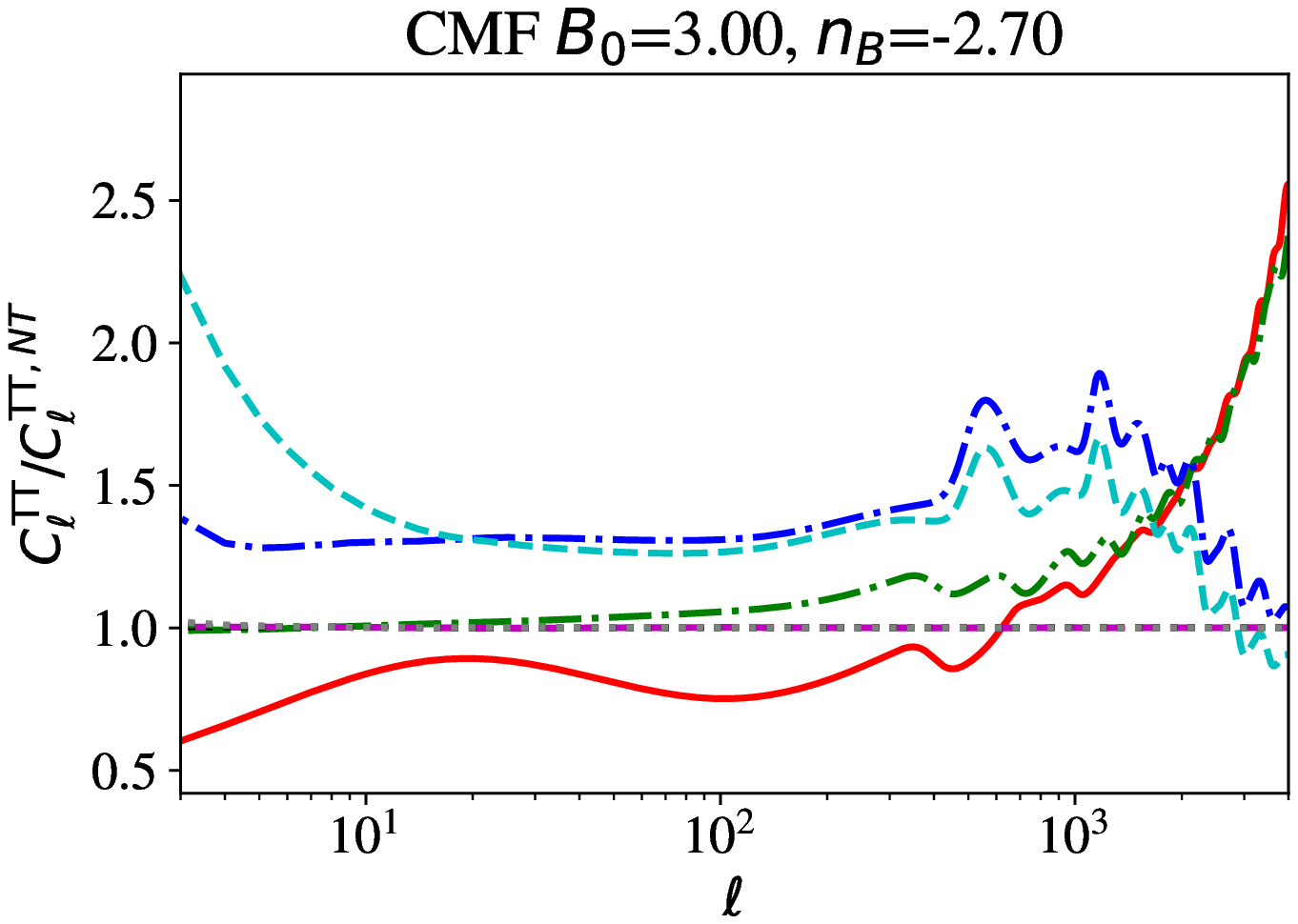}
\hspace{0.05cm}
\epsfxsize=3.2in\epsfbox{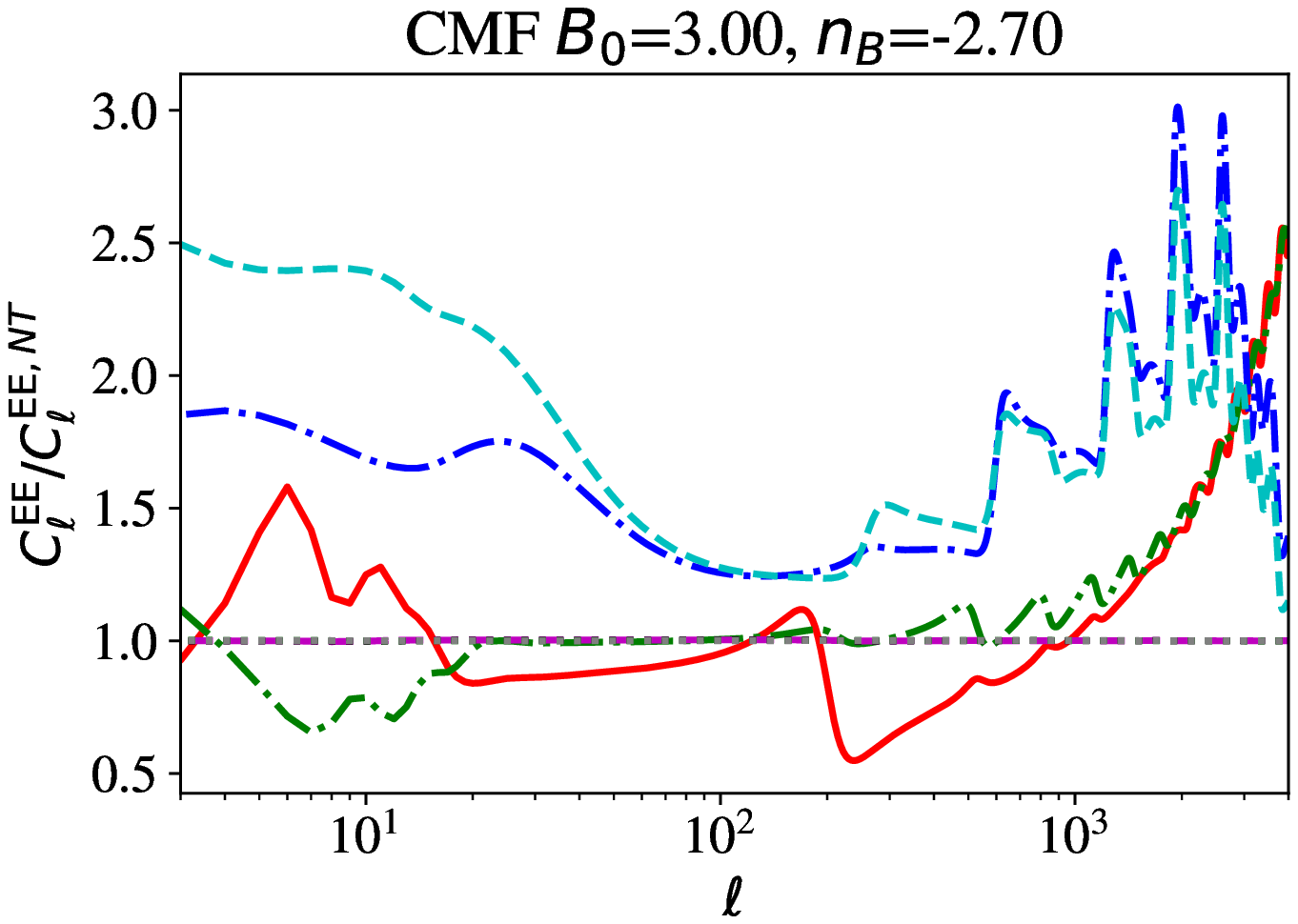}}
\centerline{
\epsfxsize=3.3in\epsfbox{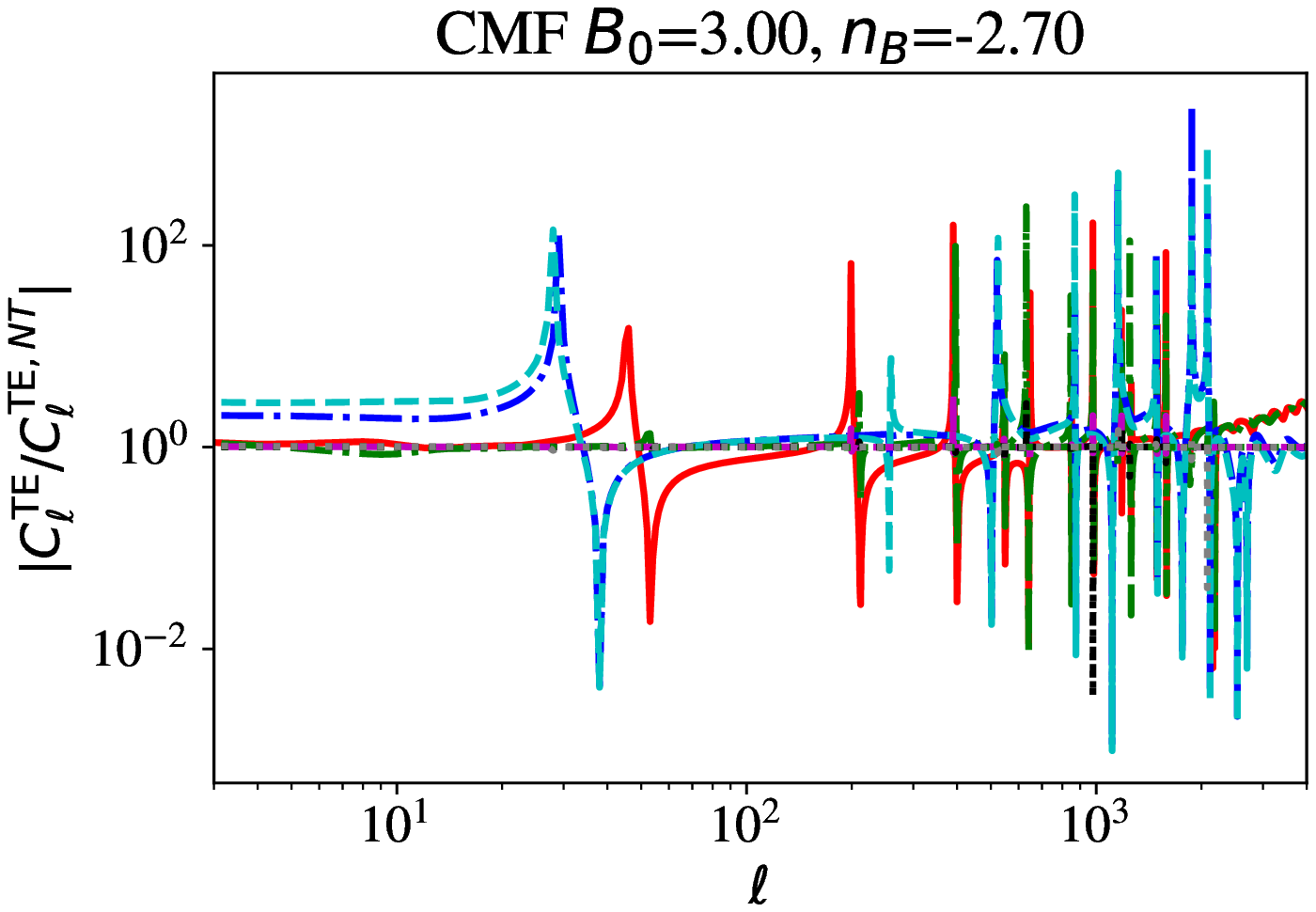}
\hspace{0.6in}
\epsfxsize=2.8in\epsfbox{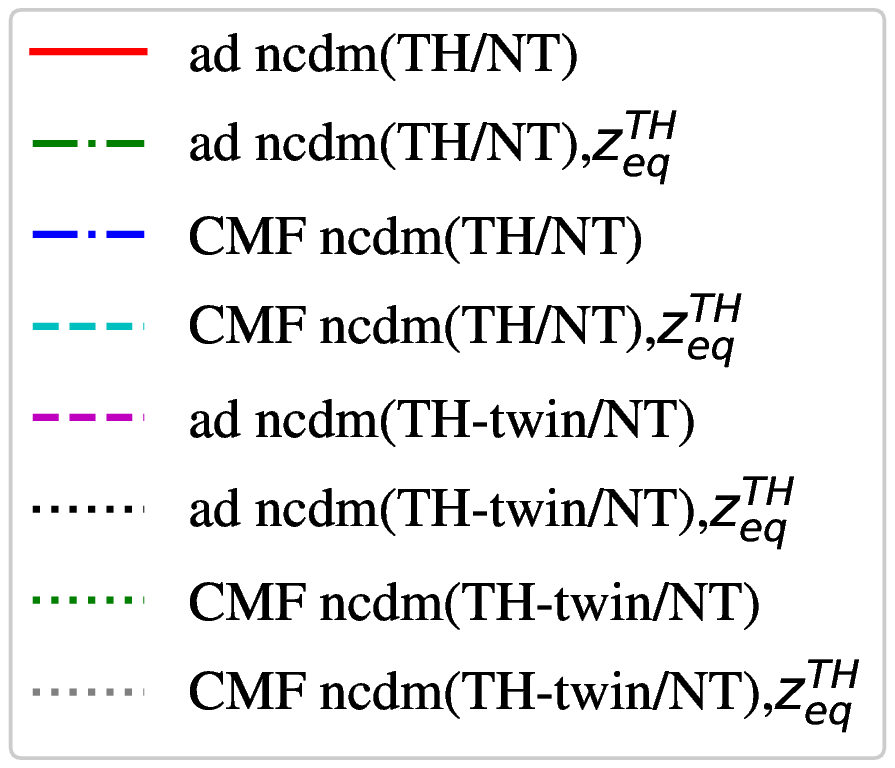}}
\caption{Ratios of the auto- and cross correlation angular power spectra of the T- and E-mode of models with three thermal neutrinos (TH), three non-thermal neutrinos (NT), three non-thermal neutrinos (NT, $z_{eq}^{TH}$) with $\omega_c$ such that radiation-matter equality occurs at the same redshift as in the three thermal neutrino model $z_{eq}^{TH}$ and their thermal twin models with extra relativistic degrees of freedom 
for the adiabatic mode (ad) and the compensated magnetic mode (CMF) with $B_0=3$ nG, $n_B=-2.70$. The curves corresponding to the ratios of the twin model of the NT model and the NT model itself  for the different modes all coincide at the horizontal line at 1.0 indicating a good numerical description of the NT models in terms of their corresponding TH twin model.
}
\label{fig3}
\end{figure}

\begin{figure}[h!]
\centerline{\epsfxsize=3.2in\epsfbox{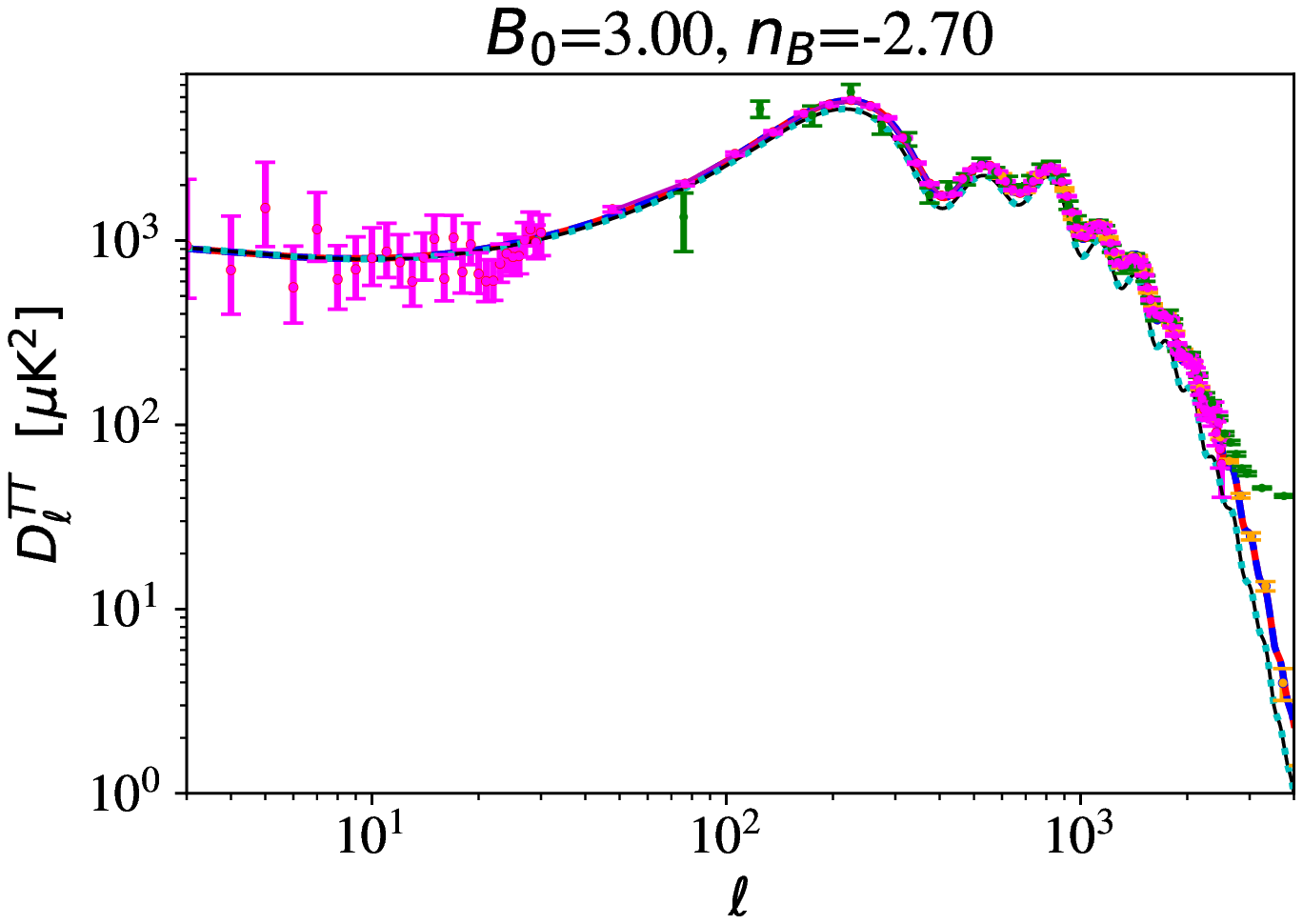}
\hspace{0.05cm}
\epsfxsize=3.2in\epsfbox{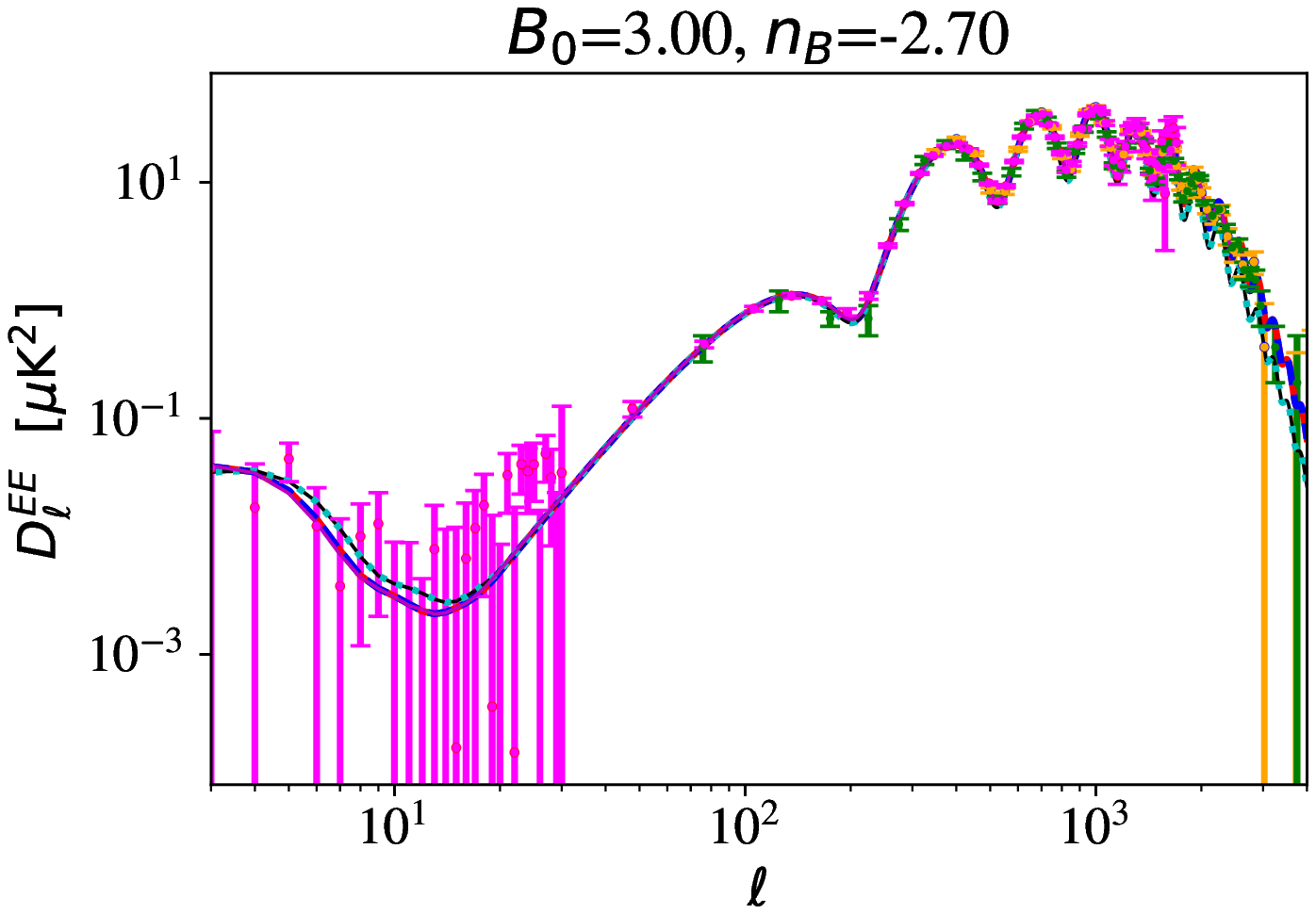}}
\centerline{
\epsfxsize=3.3in\epsfbox{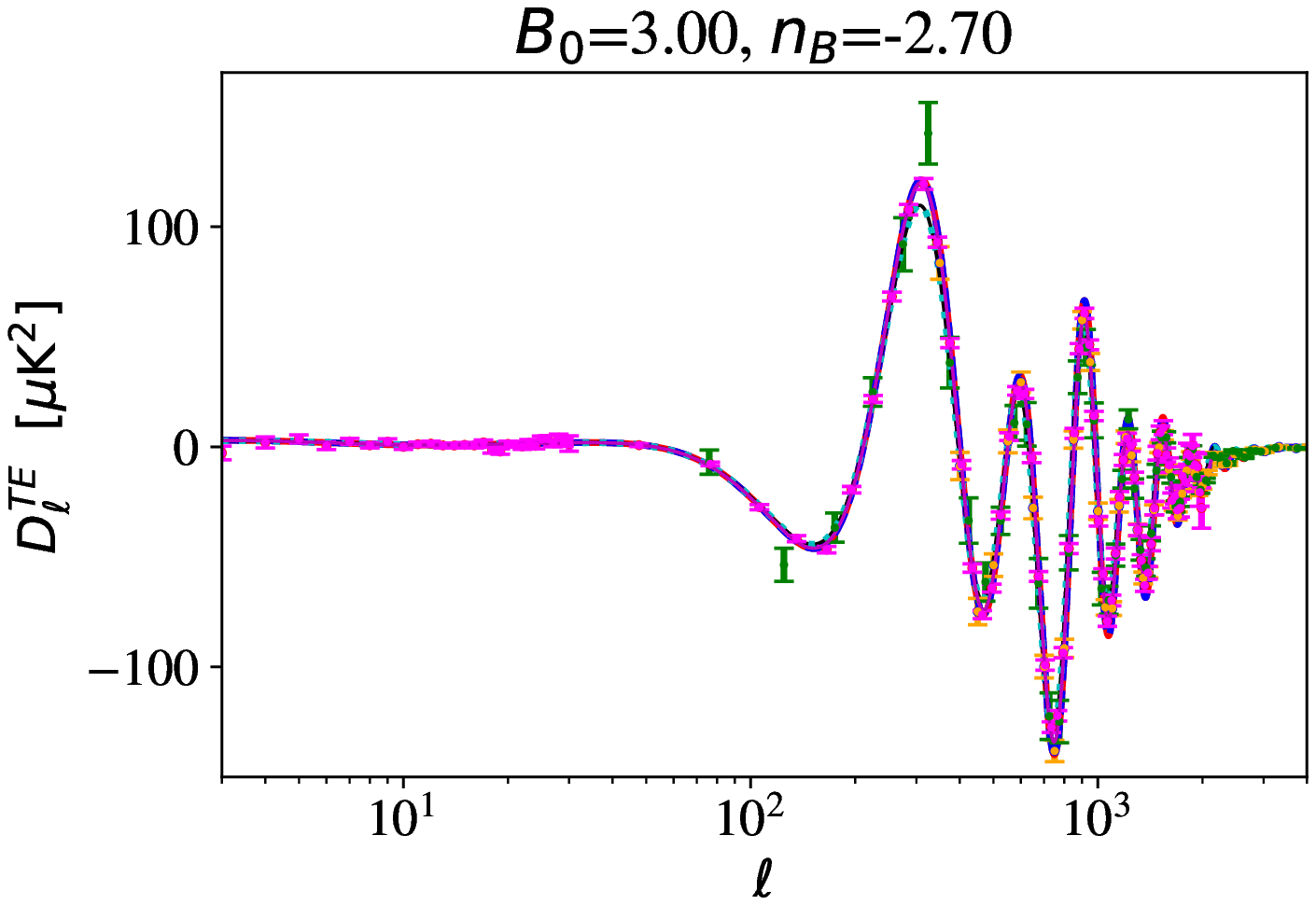}
\hspace{0.6in}\vspace{1cm}
\epsfxsize=2.8in\epsfbox{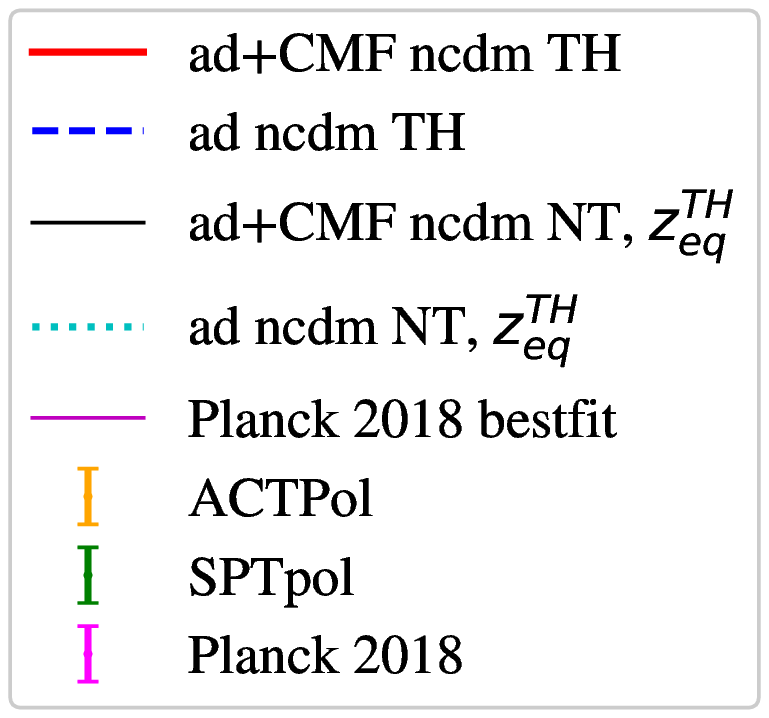}}
\caption{ Auto and cross correlation angular power spectra for $B_0=3$ nG, $n_B=-2.70$ for the thermal and non-thermal neutrino models (NT, $z_{eq}^{TH}$) 
together with data points from Planck 2018 \cite{Planck-2018}, 
ACTPol \cite{ACTPol2017} and SPTpol \cite{SPTpol2018}. Shown are the total (adiabatic plus magnetic mode) angular power spectra as well as those for only the adiabatic mode.}
\label{fig4}
\end{figure}

\begin{figure}[h!]
\centerline{\epsfxsize=3.2in\epsfbox{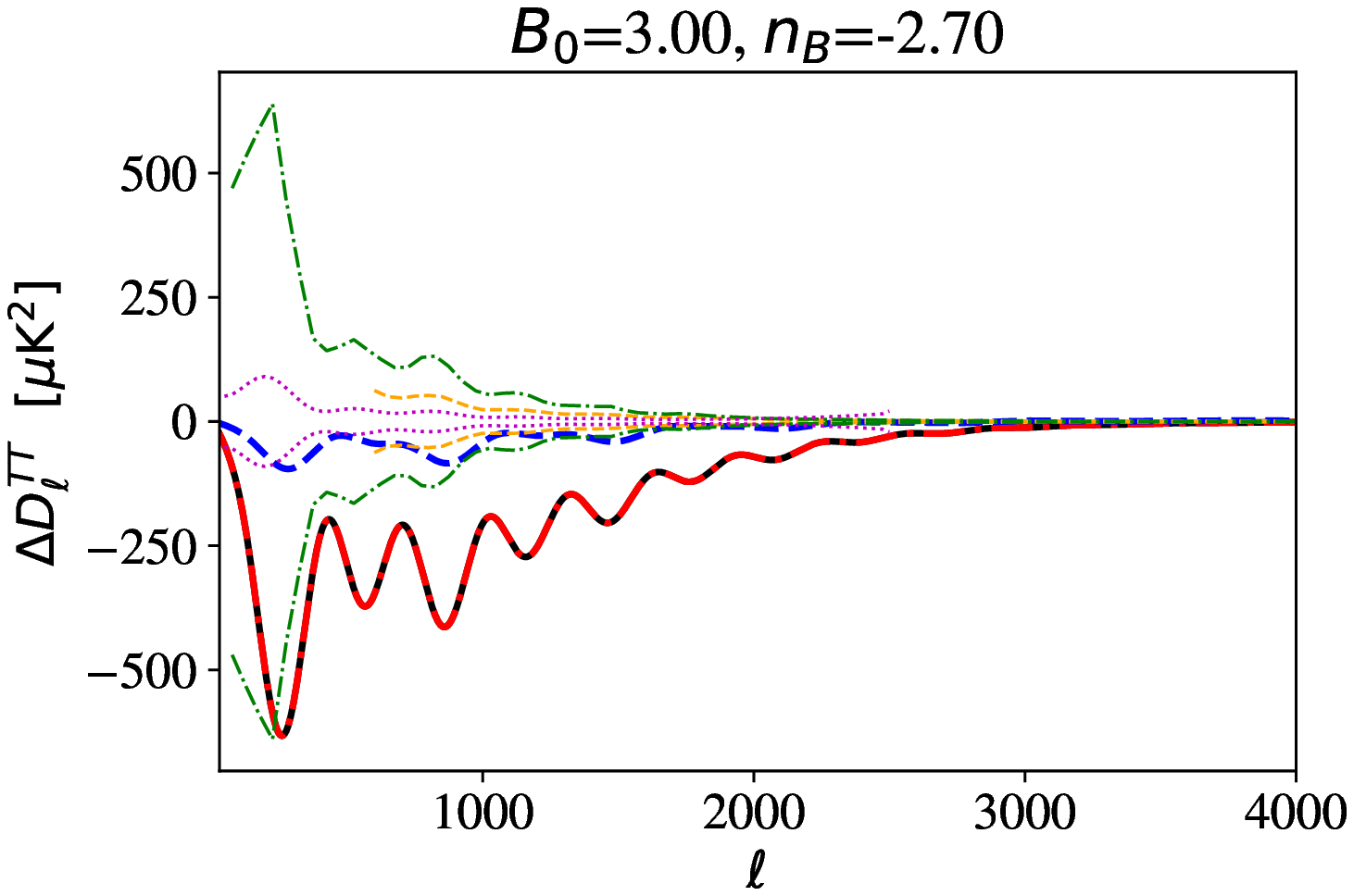}
\hspace{0.05cm}
\epsfxsize=3.1in\epsfbox{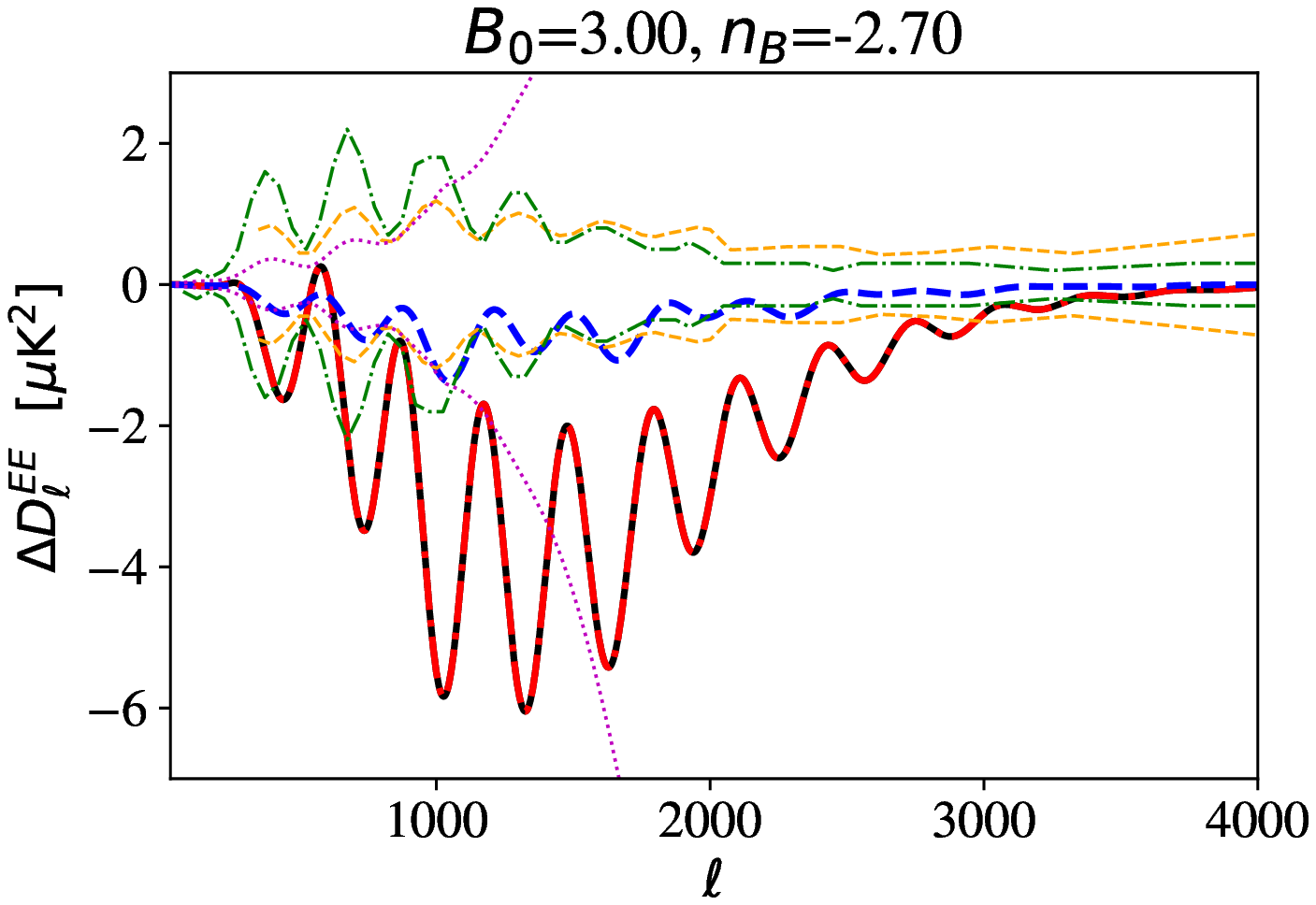}}
\centerline{
\epsfxsize=3.2in\epsfbox{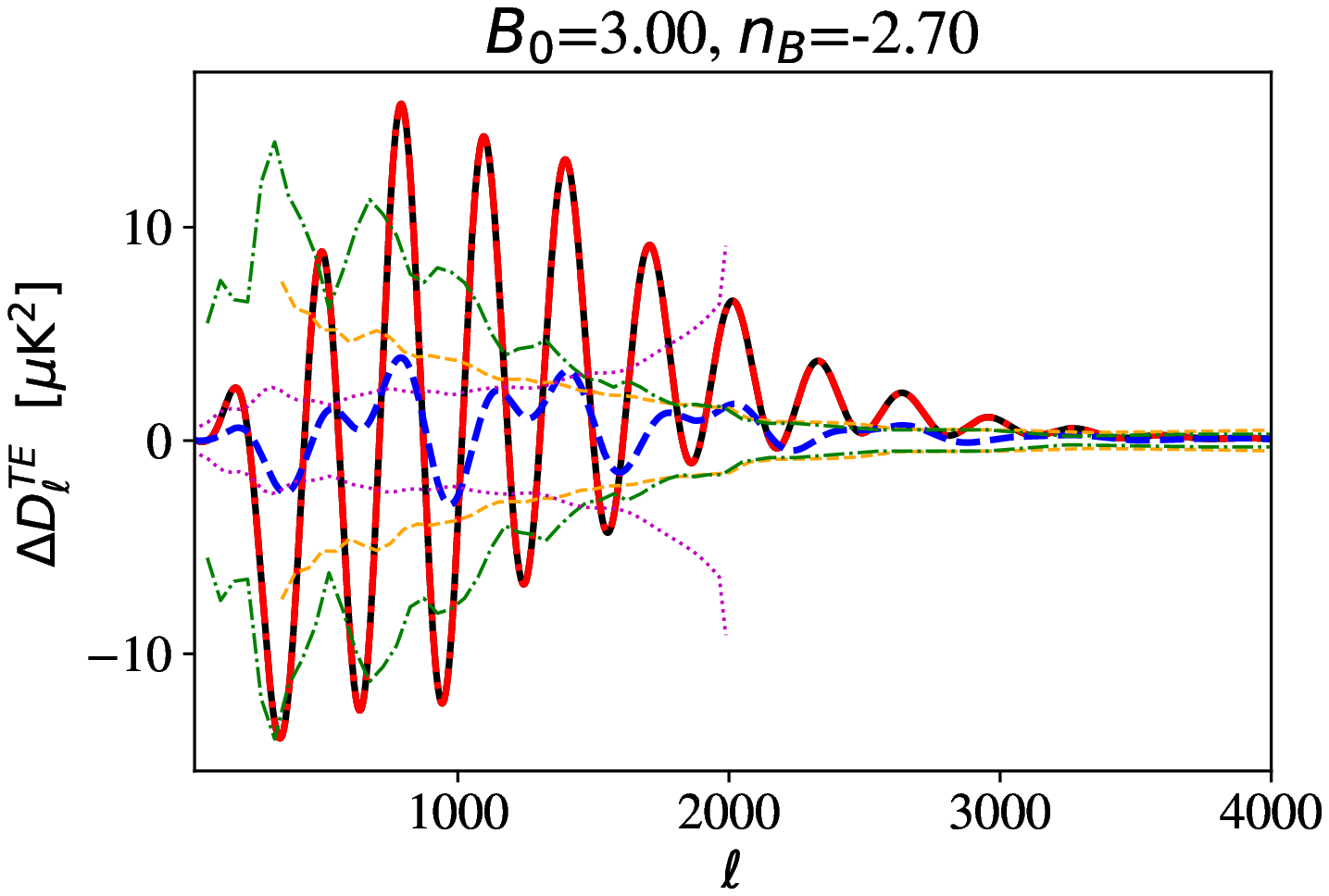}
\hspace{0.4in}
\epsfxsize=2.8in\epsfbox{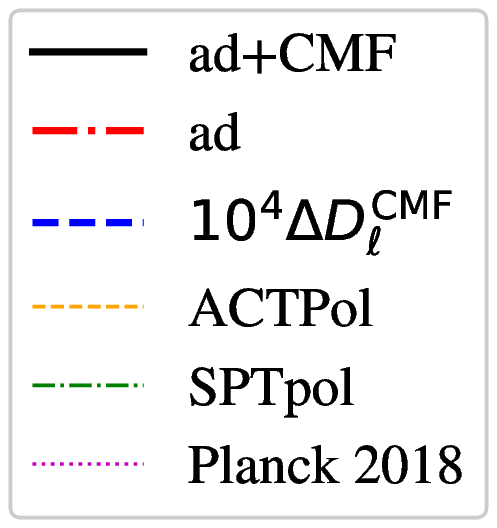}}
\caption{Differences $\Delta D_{\ell}$ (cf. equation (\ref{delD})) between non-thermal (NT, $z_{eq}^{TH})$ and thermal neutrino models in the auto and cross correlation angular power spectra for $B_0=3$ nG, $n_B=-2.70$ together  with  errors from Planck 2018 \cite{Planck-2018},  ACTPol \cite{ACTPol2017} and SPTpol \cite{SPTpol2018}. For clearness the magnetic mode $10^4 \Delta D_{\ell}^{\rm CMF}$ has been plotted.}
\label{fig5}
\end{figure}

In figure \ref{fig2} the angular power spectra of the CMB anisotropies in terms of 
\begin{eqnarray}
D_{\ell}=\frac{\ell(\ell+1)C_{\ell}}{2\pi} 
\end{eqnarray}
are shown for the three thermal neutrino model ({\it i.}), the three non-thermal neutrino model ({\it ii.})
 as well  as the corresponding twin three thermal plus extra relativistic degrees of freedom (TH+R (twin))
model ({\it iii.})
for the adiabatic, primordial curvature mode and the compensated magnetic mode for $B_0=3$ nG, $n_B=-2.7$.  
Ratios of the angular power spectra of models with thermal and non-thermal neutrinos, respectively, are shown in figure \ref{fig3}.
As can be appreciated from the horizontal lines at 1.0 in figure \ref{fig3}  the three non-thermal neutrino model ({\it ii.}) is well described by the  three thermal neutrinos plus extra relativistic degrees of freedom model ({\it iii.}). This holds for the adiabatic mode as well as the compensated magnetic mode.
In figure \ref{fig4} for the same magnetic field parameters as in figures \ref{fig2} and \ref{fig3}  the thermal and non-thermal neutrino models (NT, $z_{eq}^{TH}$) for the total angular power spectra with contributions from the adiabatic as well as the compensated magnetic mode are shown together with data points from Planck 2018 \cite{Planck-2018}, 
ACTPol \cite{ACTPol2017} and SPTpol \cite{SPTpol2018}.
As can be appreciated from figure \ref{fig4} the dominant contribution to the total angular power spectra comes from the adiabatic mode which can be seen explicitly in  figure \ref{fig2}. The visible difference is between the thermal and non-thermal neutrino models. Whereas the thermal neutrino model fits the data very well this is not
the case for the non-thermal neutrino model for all multipoles. This can also be clearly seen in figure \ref{fig5} where 
the difference is shown between the  angular power spectra of the non-thermal and thermal neutrino models 
\begin{eqnarray}
\Delta D_{\ell}=D_{\ell}^{(NT,z_{eq}^{TH})}-D_{\ell}^{(TH)}
\label{delD}
\end{eqnarray}
together with  errors  from Planck 2018 \cite{Planck-2018}, 
ACTPol \cite{ACTPol2017} and SPTpol \cite{SPTpol2018}.
This result is not surprising as $N^{NT}_{eff}=8.049$ for the choice of model parameters 
of the three non-thermal (NT) neutrino model ({\it ii.)} used here as  way of example to study in particular the effect on the compensated magnetic mode, as pointed out above.

\begin{figure}[h!]
\centerline{\epsfxsize=3.2in\epsfbox{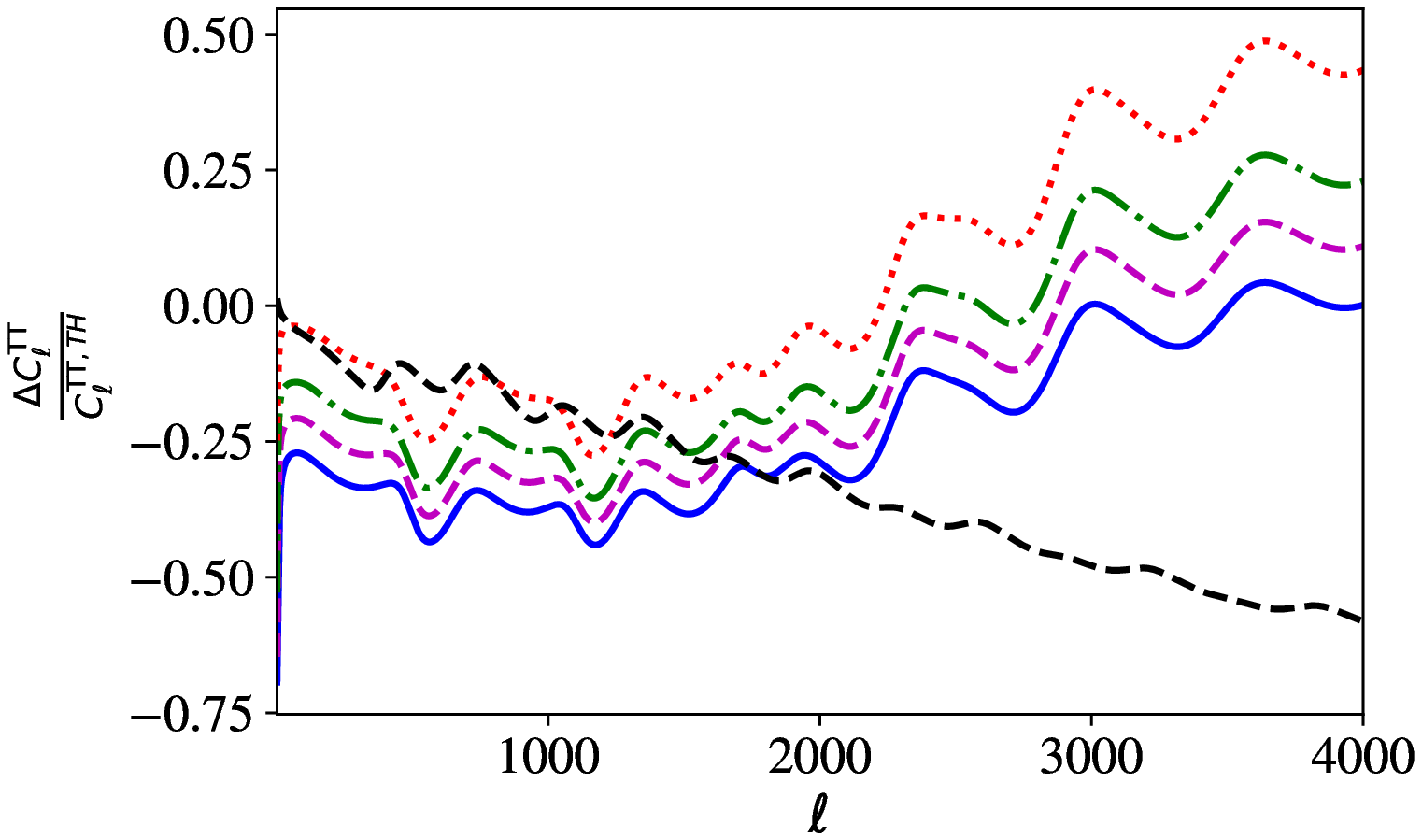}
\hspace{0.05cm}
\epsfxsize=3.2in\epsfbox{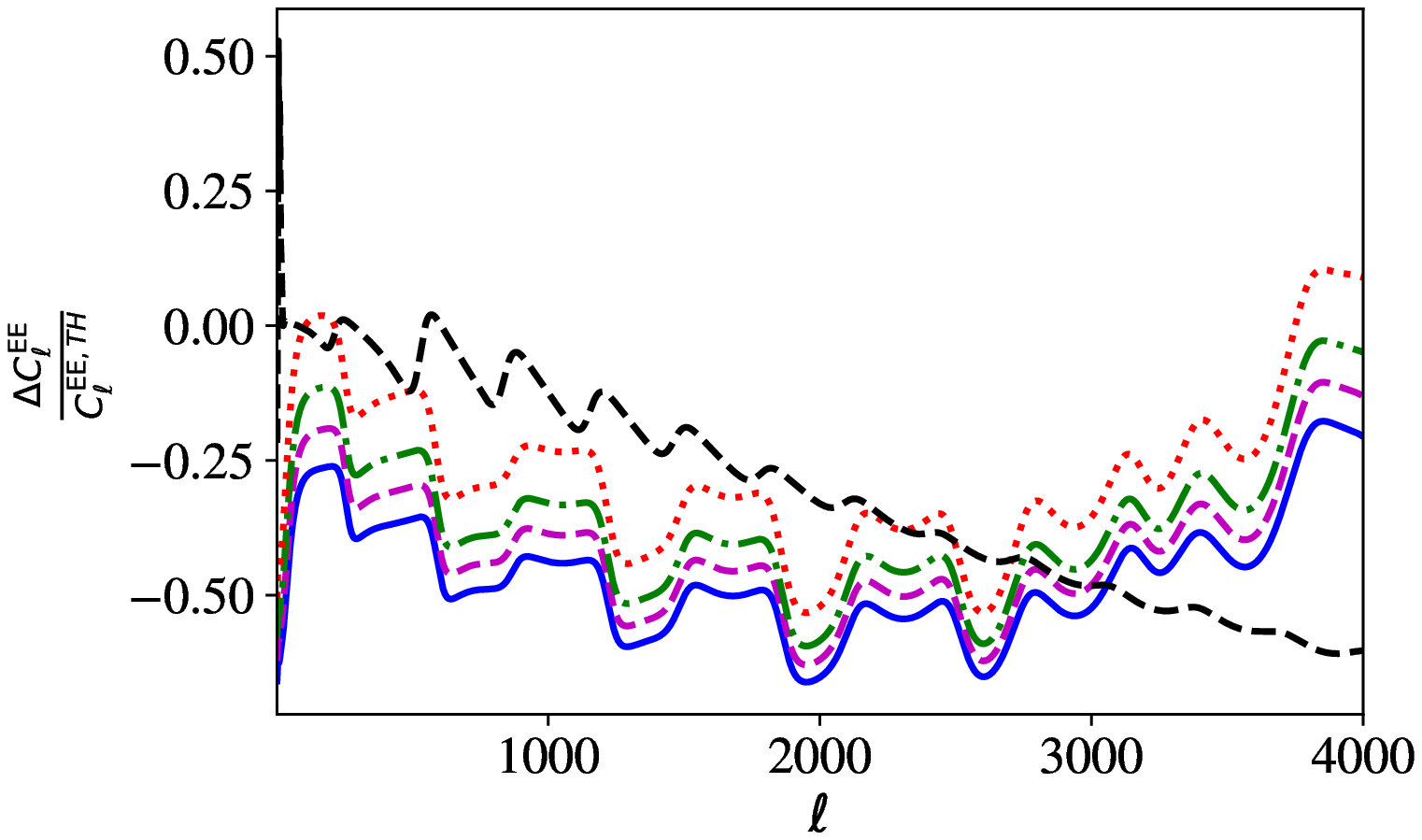}}
\centerline{
\epsfxsize=3.1in\epsfbox{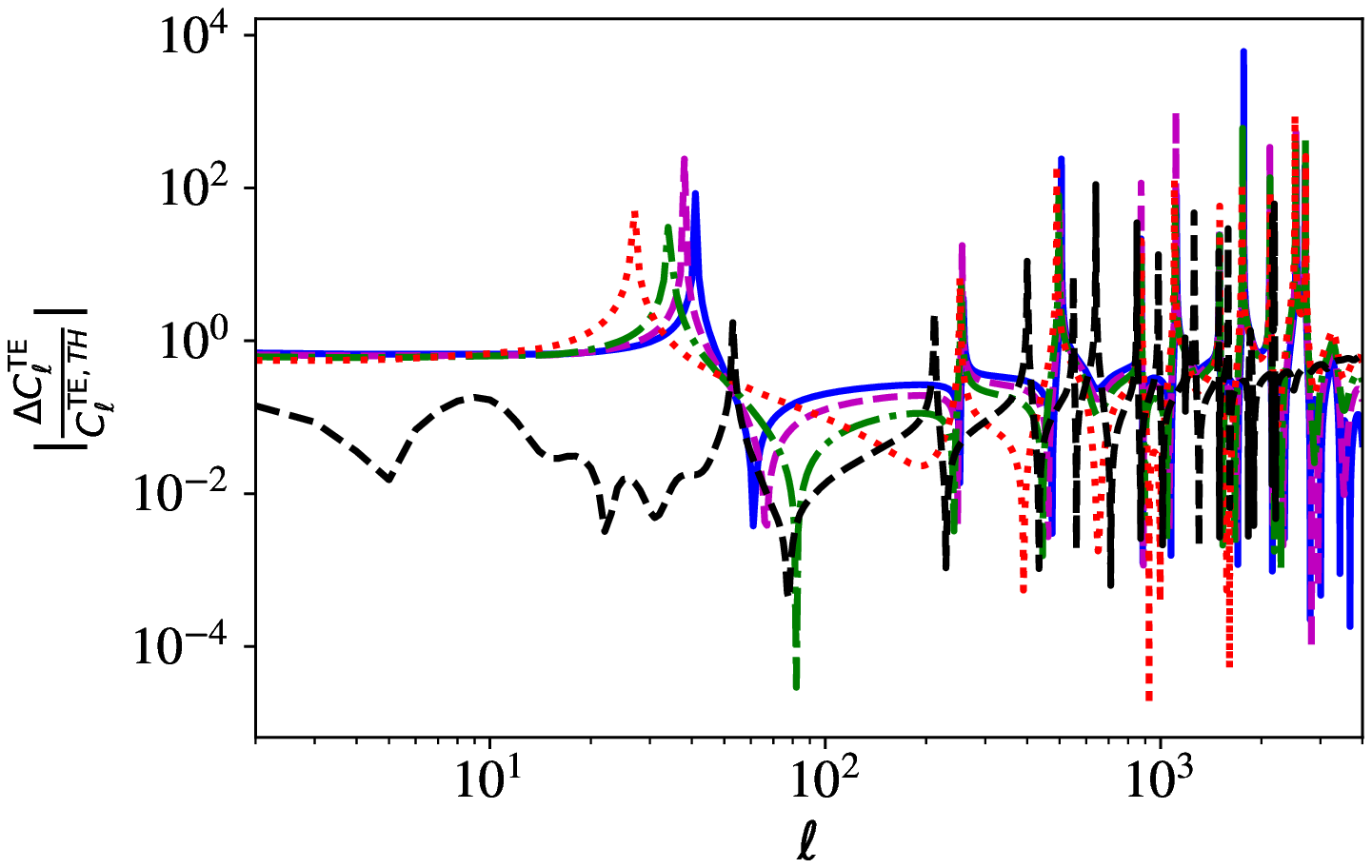}
\hspace{0.6in}\vspace{1cm}
\epsfxsize=2.8in\epsfbox{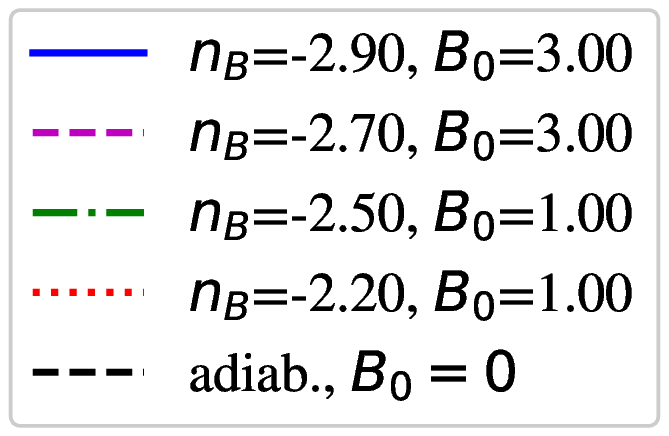}}
\caption{Relative change of the CMB angular power spectra of the  three non-thermal neutrinos (NT, $z_{eq}^{TH}$)  w.r.t. the three thermal neutrinos model (TH) (cf. equation
(\ref{DelMoM}))
for different choices of the magnetic field parameters for the compensated magnetic mode ($B_0$[nG], $n_B$) as well as the adiabatic, primordial curvature mode.
}
\label{fig6}
\end{figure}

To compare power spectra for the thermal and non-thermal cases it is useful to define the corresponding relative change by
\begin{eqnarray}
\frac{\Delta M}{M}\equiv\frac{M^{(NT, z^{TH}_{eq})}-M^{(TH)}}{M^{(TH)}}
\label{DelMoM}
\end{eqnarray}
where in the following $M$ denotes the angular power spectra of the T-mode and of  the E-mode auto correlation, $C_{\ell}^{TT}$ and $C_{\ell}^{EE}$, respectively, as well as the linear matter power spectrum $P(k)$.
In figure \ref{fig6} the relative change of the CMB angular power spectra, $\frac{\Delta C_{\ell}}{C_{\ell}}$ of the adiabatic as well as the 
compensated magnetic mode are shown for the T-mode and 
E-mode auto- and cross correlations for a set of different magnetic field parameters.

\begin{figure}[h!]
\centerline{
\epsfxsize=3.8in\epsfbox{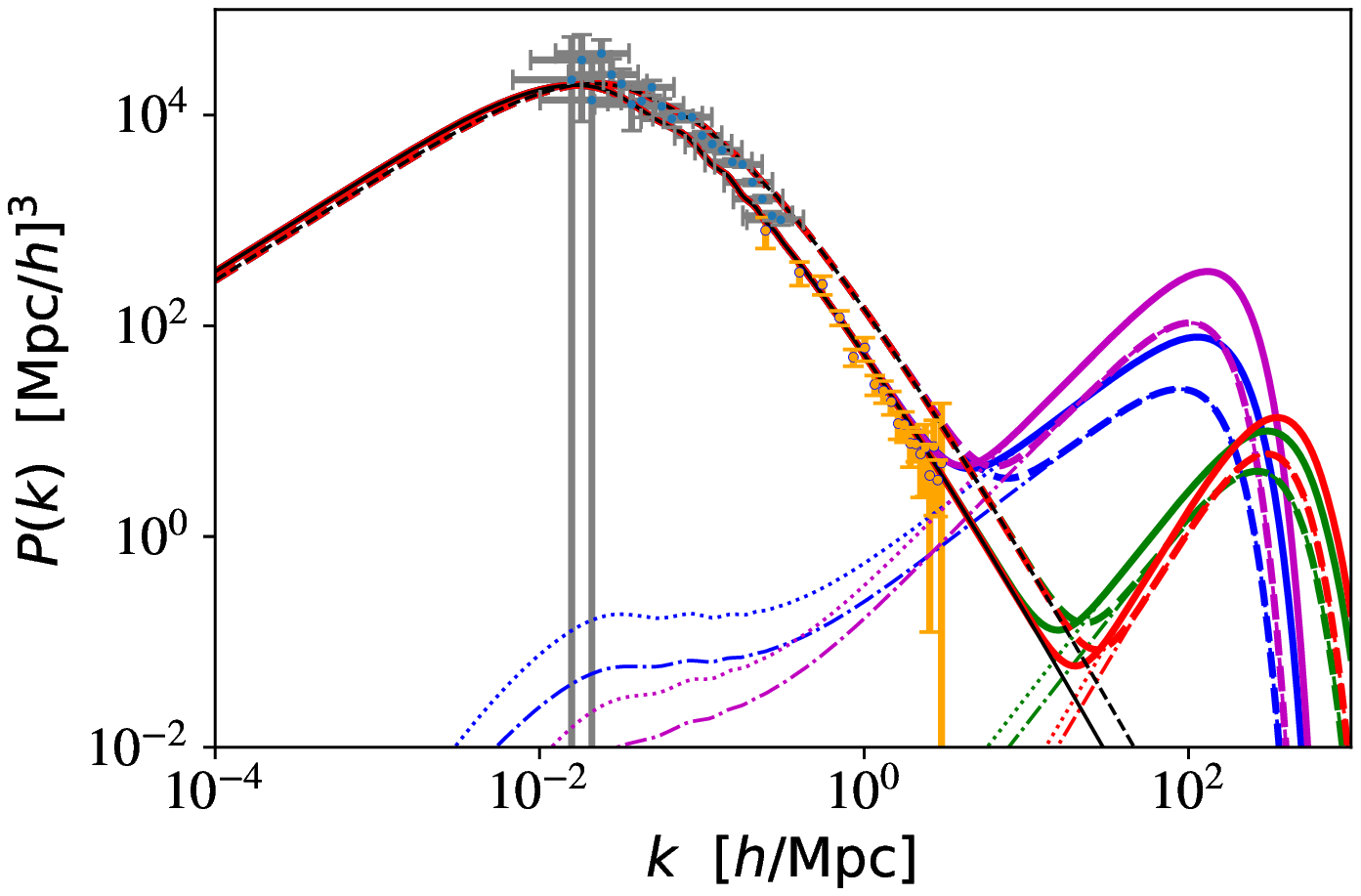}
\hspace{0.6cm}
\epsfxsize=2.8in\epsfbox{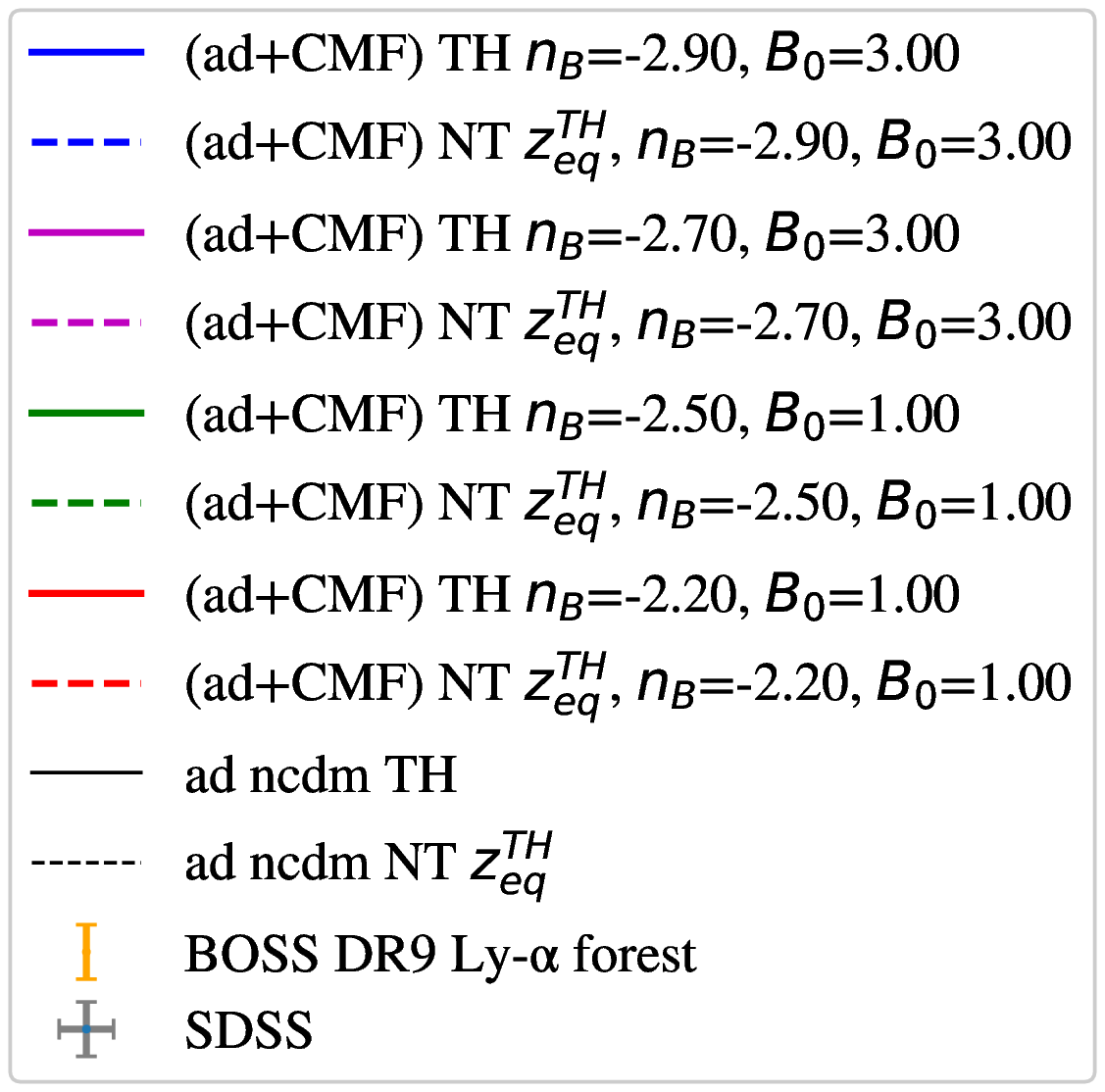}}
\centerline{
\epsfxsize=3.8in\epsfbox{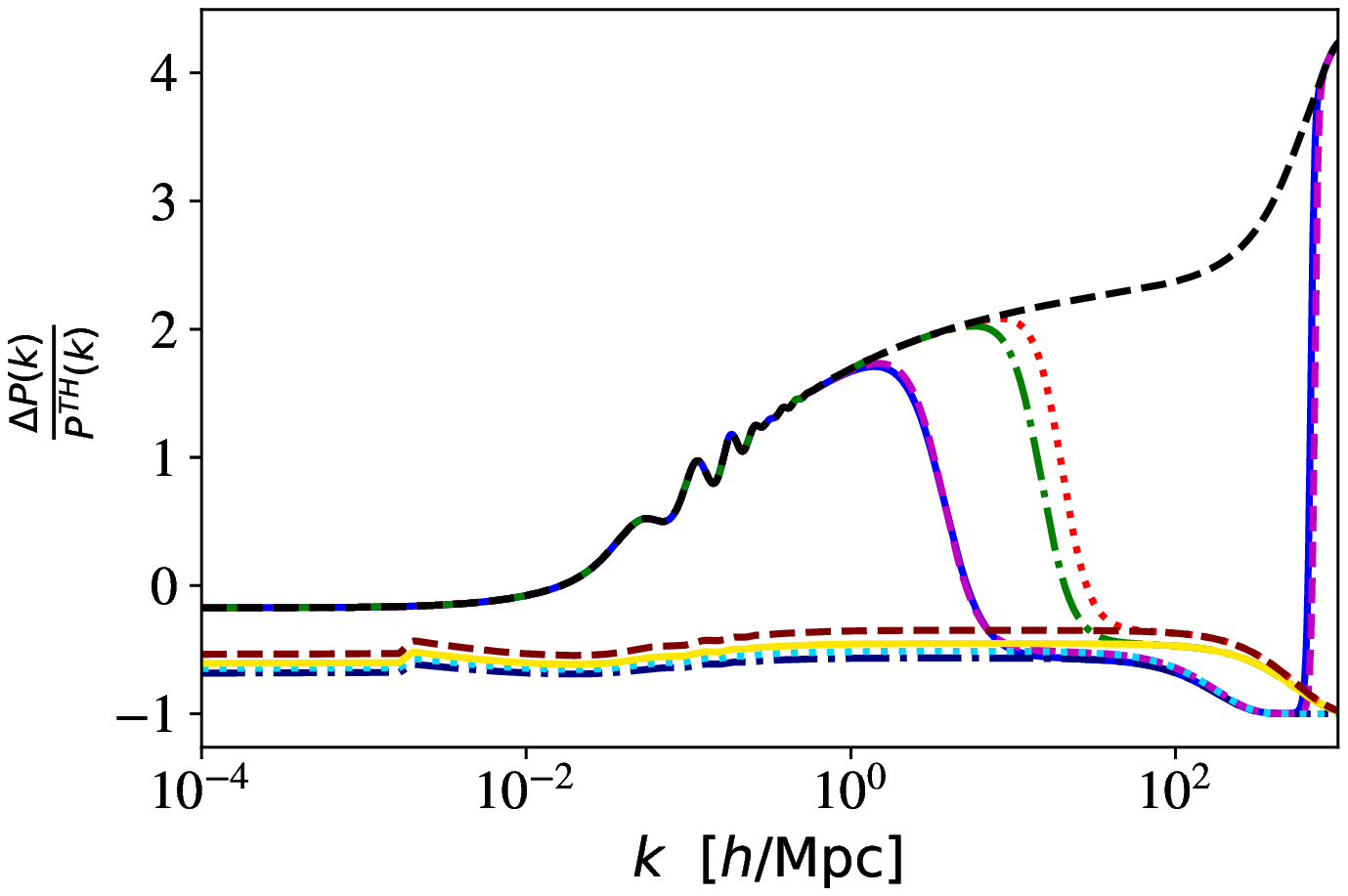}
\hspace{0.7cm}
\epsfxsize=2.8in\epsfbox{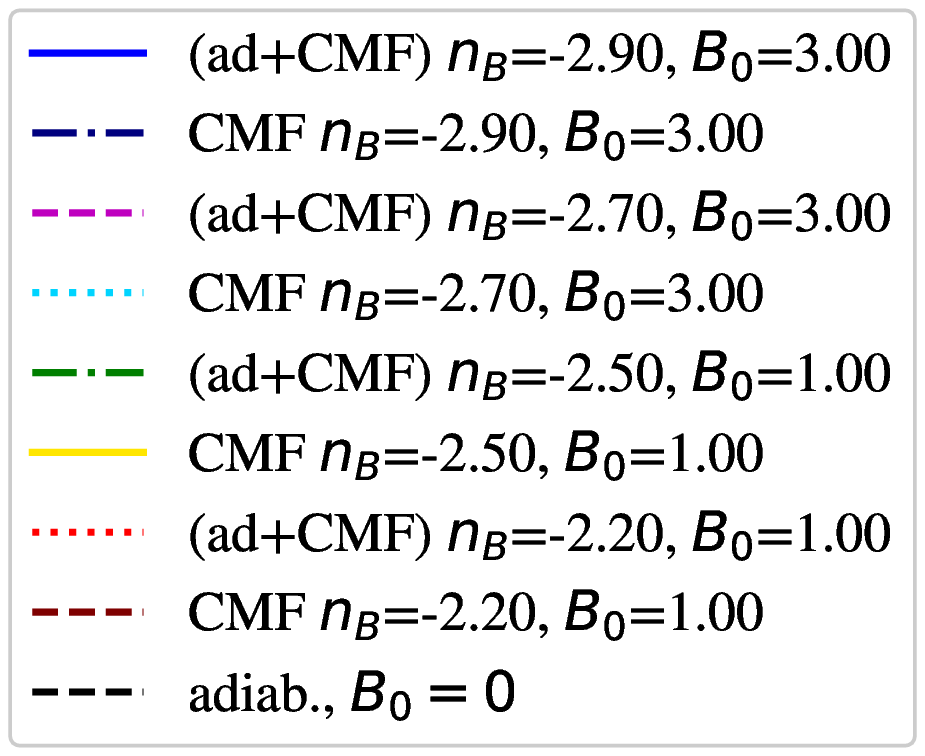}}
\caption{Linear matter power spectrum for three thermal neutrinos (TH) and three non-thermal neutrinos (NT, $z_{eq}^{TH}$) with distribution function (\ref{pdf}) for different choices of the magnetic field parameters ($B_0$[nG], $n_B$). {\sl Upper panel:} The total linear matter power spectrum of the adiabatic mode (ad) and the compensated magnetic mode (CMF) is shown together with data points from BOSS DR9 Ly-$\alpha$ forrest \cite{bossLyalpha} and SDSS \cite{sdss}. The light dotted and dashed-dottted lines indicate the three neutrino thermal magnetic mode and 
the three neutrino non-thermal pure magnetic mode solutions, respectively. $z_{eq}^{TH}$ denotes that the cosmological parameters have been adjusted so that the redshift of radiation-matter equality in the non-thermal model is the same as that in the non-thermal one (see details in the text).
{\sl Lower panel:} Relative change of the linear matter power spectrum w.r.t to the three thermal neutrino model (TH).
}
\label{fig7}
\end{figure}

In figure \ref{fig7}  the linear matter power spectra as well as the relative changes between the thermal and non-thermal neutrino models are shown for different values of the magnetic field parameters. 
In the presence of a magnetic field the magnetic Jeans length, corresponding to a wave number $k_{mJ}$,  is a characteristic scale at which magnetic pressure support prevents gravitational collapse. However, as pointed out in \cite{KimOlinRos} the  density perturbation spectrum  is cut-off at  $k_{mJ}$ only in a purely baryonic universe. As cold dark matter does not couple to the magnetic field the total matter power spectrum is flattened out but not cut-off at $k_{mJ}$. To model this correctly would require to include  magnetohydrodynamical non linear effects which is beyond the scope of this paper.

As can be seen in figure \ref{fig7} on small scales the effect of the Lorentz term is clearly visible. During the matter dominated era on subhorizon scales the linear matter power spectrum can be approximated by \cite{kk14}
\begin{eqnarray}
P_m^{(B)}=\frac{2\pi}{k^3}\left(\frac{k}{a_0H_0}\right)^4\frac{4}{225}
(1+z_{dec})^2\left(\frac{\Omega_{\gamma,0}}{\Omega_{m,0}}\right)^2
{\cal P}_L(k)
\end{eqnarray}
where ${\cal  P}_L(k)$ is the dimensionless power spectrum of the Lorentz term (cf. equation (\ref{L})).
Keeping the redshift of radiation-matter equality fixed imposes (e.g., \cite{lmmp})
\begin{eqnarray}
\Omega_{m,0}=\Omega_{\gamma,0}\left[1+\frac{7}{8}\left(\frac{4}{11}\right)^{\frac{4}{3}}N_{eff}^{NT,TH-twin}
\right](1+z_{eq}^{TH}),
\end{eqnarray}
where the equivalence of the three non-thermal neutrino model in terms of the three thermal
neutrino plus extra relativistic degrees of freedom ((TH+R) (twin)) (cf. figures \ref{fig2} and \ref{fig3}) is used.
Since the three  non-thermal neutrino model corresponds to  additional relativistic degrees of freedom 
the effective coupling of the Lorentz term in the matter power spectrum is reduced. This can be observed in the 
numerical solutions in figure \ref{fig7}. In particular, the relative change w.r.t. the three thermal neutrino model in the linear matter power spectrum is negative
as can be seen in figure \ref{fig7} ({\sl lower panel}).
In figure \ref{fig7} ({\sl upper panel}) it can be appreciated that on larger scales the total linear matter power spectrum is dominated by the contribution of the adiabatic, primordial 
curvature mode. Thus its amplitude is enhanced in the three non-thermal neutrino model ({\it ii.)}.
On the contrary, 
on small scales the contribution of the compensated magnetic mode dominates over that of the adiabatic mode
leading to a suppression of the total  linear matter power spectrum in the three non-thermal neutrino model ({\it ii.)}.

Moreover, there seem to be indications for a new degeneracy with  the parameters of the magnetic field. 
For example, in figure \ref{fig7} the numerical solutions for the magnetic field 
with $B_0=3$ nG and for  $n_B=-2.9$ the three neutrino thermal model ({\it i.)} and for $n_B=-2.7$ the three non-thermal neutrino model ({\it ii.}) are quite close. 
Thus in this case changing the magnetic field
spectral index leads to a  numerical solution of a three non-thermal neutrino model which effectively corresponds to that of a three thermal neutrino model.

In figure \ref{fig7} ({\sl upper panel}) data points from  BOSS DR9 Ly-$\alpha$ forrest \cite{bossLyalpha} and SDSS \cite{sdss} have been included. 
The total linear matter power spectrum of the adiabatic, primordial curvature mode and the compensated magnetic mode for the numerical examples of the three neutrino non-thermal model just fit the error bars of the SDSS data points and are excluded by most of the BOSS data points. However, it should be kept in mind that the particular form of the
non-thermal neutrino phase-space distribution function as well as the numerical values of the model parameters have been chosen to study the effects in general. A detailed parameter estimation study is left to future work.

\section{Conclusions}
\label{s3}
\setcounter{equation}{0}

Models with three thermal light neutrinos and three non-thermal light neutrinos with the same degenerate mass configuration have been studied in the presence of a primordial stochastic magnetic field. The non-thermal neutrino distribution function is modelled by a Fermi-Dirac distribution with an additional Gaussian peak. This type of distribution could be the result of a scalar particle decaying into neutrinos after decoupling of the neutrinos of the standard model of cosmology \cite{Cuoco:2005}. 
The numerical solutions for the angular power spectra of the CMB anisotropies as well as the linear matter power spectrum have been found by modifying the CLASS code accordingly.
There is a known  degeneracy between (light) neutrino masses $m_0$, the cold dark matter density parameter $\omega_{c}$ and the number of relativistic degrees of freedom $N_{eff}$ for $\Lambda$CDM models leading to an equivalent, effective description of models including non-thermal neutrinos in terms of a twin thermal neutrino model with extra relativistic degrees of freedom \cite{Cuoco:2005}. Here it has been found that this effective description can be extended to models with non-thermal neutrinos in the presence of a stochastic magnetic field. Moreover, an additional degeneracy with the magnetic field parameters has been observed in the numerical solutions allowing to connect
for the same magnetic field amplitude a three thermal neutrino model with a three non-thermal neutrino model by changing the magnetic field spectral index.

Moreover, it is found that the amplitude of the  linear matter power spectrum of the three neutrino non-thermal pure compensated magnetic mode is suppressed in comparison to the one in the three neutrino thermal pure compensated magnetic mode model.
This is the opposite behaviour of the adiabatic, primordial curvature mode where the amplitude is larger in the case of the three non-thermal neutrino model.
The suppression of the matter perturbation of the three non-thermal neutrino compensated magnetic mode is related to the diminished coupling to the Lorentz term because of a larger cold dark matter density parameter. Thus magnetic field spectra with larger amplitudes or stronger tilt can be compensated by light neutrinos with a non-thermal phase space distribution.

\section{Acknowledgments}

The use of the Planck Legacy Archive \footnote{https://pla.esac.esa.int/pla/\#cosmology} is gratefully acknowledged as well  as the 
use of the Legacy Archive for Microwave Background Data Analysis (LAMBDA) \footnote{https://lambda.gsfc.nasa.gov}, 
part of the High Energy Astrophysics Science Archive Center (HEASARC). HEASARC/LAMBDA is a service of the Astrophysics Science Division at the NASA Goddard Space Flight Center. 
Financial support by Spanish Science Ministry grant PGC2018-094626-B-C22 (MCIU/AEI/FEDER, EU) and Basque Government grant IT979-16
is gratefully acknowledged.


\bibliography{references}

\bibliographystyle{apsrev}

\end{document}